\documentclass[pre,reprint,nobibnotes,aps,showpacs]{revtex4-1}

\usepackage[english]{babel}
\usepackage{grffile,verbatim}
\usepackage{graphicx,epstopdf,color}
\usepackage{amsmath,amssymb,psfrag}
\usepackage[caption=false]{subfig}

\usepackage[latin1]{inputenc}

\newcommand{\der}[2]{\frac{\partial #1}{\partial #2}}
\newcommand{\Pin}{P_{\mathrm{in}}}
\newcommand{\Pout}{P_{\mathrm{out}}}

\newcommand{\cATP}{[\text{ATP}]}
\newcommand{\cADP}{[\text{ADP}]}
\newcommand{\cP}{[\text{P$_{\mathrm{i}}$}]}
\newcommand{\Keq}{K_\text{eq}}

\newcommand{\vld}{v_\text{LD}}

\newcommand{\fhdld}{f_{\text{HD}-\text{LD}}}
\newcommand{\jpbc}{J_{\text{PBC}}}
\newcommand{\jlkpbc}{J^{\text{LK}}_{\text{PBC}}}
\newcommand{\jmlk}{J^{\text{m}}_{\text{LK}}}
\newcommand{\jclk}{J^{\text{c}}_{\text{LK}}}
\newcommand{\plk}{\rho_{\text{LK}}}
\newcommand{\pobc}{\rho_{\text{OBC}}}
\newcommand{\etalk}{\eta_{\text{LK}}}

\newcommand{\etaobc}{\eta_{\text{OBC}}}
\newcommand{\fobc}{f_{\text{OBC}}}
\newcommand{\vlk}{v_{\text{LK}}}
\newcommand{\rlk}{r_{\text{LK}}}
\newcommand{\vobc}{v_{\text{OBC}}}
\newcommand{\robc}{r_{\text{OBC}}}
\newcommand{\vsII}{v_0^{\text{II}}}
\newcommand{\vsI}{v_0^{\text{I}}}
\newcommand{\jmpbc}{J^{\text{m}}_{\text{PBC}}}
\newcommand{\jmcpbc}{J^{\text{m,c}}_{\text{PBC}}}
\newcommand{\jcpbc}{J^{\text{c}}_{\text{PBC}}}
\newcommand{\vpbc}{v_{\text{PBC}}}
\newcommand{\rpbc}{r_{\text{PBC}}}
\newcommand{\pld}{\rho_{\text{LD}}}
\newcommand{\phd}{\rho_{\text{HD}}}
\newcommand{\pmc}{\rho_{\text{MC}}}
\newcommand{\prmax}{\rho_r^{\text{max}}}
\newcommand{\plmin}{\rho_l^{\text{min}}}
\newcommand{\jm}{J^{\text{m}}}

\newcommand{\win}{w_{\text{in}}}
\newcommand{\wout}{w_{\text{out}}}

\newcommand{\phos}{\text{P}_{\text{i}}}

\definecolor{green}{rgb}{0,.3922,0}

\usepackage{siunitx}
\DeclareSIUnit\Molar{\textsc{m}}

\begin{document}

\title{Maximum power operation of interacting molecular motors}
\author{N. Golubeva}
\author{A. Imparato}
\affiliation{Department of Physics and Astronomy, University of Aarhus, Ny Munkegade, Building 1520, DK--8000 Aarhus C, Denmark}
\date{January, 2013}

\begin{abstract}
We study the mechanical and thermodynamic properties of different traffic models for kinesin which are relevant in biological and  experimental contexts. We find that motor-motor interactions play a fundamental role by enhancing the thermodynamic efficiency at maximum power of the motors, as compared to the non-interacting system, in a wide range of biologically compatible scenarios. We furthermore consider the case where the motor-motor interaction directly affects the internal chemical cycle and investigate the effect on the system dynamics and thermodynamics.
\end{abstract}

\pacs{05.70.Ln, 05.40.-a, 87.16.Nn}
\maketitle

\section{Introduction}
Molecular motors are biological machines that harness chemical energy and convert it into motion or useful mechanical work. These molecular machines are responsible for performing tasks as diverse as DNA replication and repair, RNA transcription, protein synthesis and intracellular transport \cite{Alberts2007}. The development of sophisticated single-molecule experimental techniques and the emergence of several theoretical frameworks for modelling single motors has over the past two decades led to a great amount of accumulated knowledge on mechanics and thermodynamics of single molecular machines \cite{Julicher1997,Kolomeisky2007,Seifert2011,Verley2011}. However, many motors, such as, e.g., kinesin or dynein motors involved in cellular cargo transport, move on crowded filamenteous tracks known as microtubules where they can encounter other motors. Such encounters give rise to non-negligible motor-motor interactions and affect the resulting motion of the motors. Molecular motor traffic is therefore an important and widely studied phenomenon, which is typically modelled by using exclusion processes on lattices \cite{Aghababaie1999,Slanina2008,Klumpp2005,Mueller2008,Nishinari2005,Brugues2009,Klumpp2008a,Ciandrini2010,Neri2011,Parmeggiani2003,ASEP}.
     
In this paper we extend and study in detail two different traffic models for kinesin, which were first used in \cite{Golubeva2012a} to study the efficiency of kinesin operating under external mechanical load force in the maximum power regime. Model I represents the simplest possible description of molecular motor traffic which is obtained by neglecting the internal conformational states of the molecular motor and modelling the system as a standard asymmetric simple exclusion process (ASEP) \cite{ASEP}. In this framework, the motor dynamics is represented as a continuous-time, stochastic jumping process of particles on a discrete lattice. Kinesin is powered by the ATP hydrolysis reaction, in which an adenosine triphosphate (ATP) molecule is hydrolyzed into an adenosine diphosphate (ADP) and a phosphate ($\phos$) molecule. The forward stepping is thus associated with ATP hydrolysis, while the backward jumps must proceed through ATP synthesis in this minimal model in order to obtain a thermodynamically consistent description.
The particle stepping is subject to an exclusion rule; if the particle attempts to step either forward or backward to a neighbouring site that is already occupied by another motor, the step will be rejected. The ASEP and extensions hereof have been used in several studies of molecular motor traffic \cite{Klumpp2005,Mueller2008,Parmeggiani2003}. 

However, all these models do not take into account that the stepping of a molecular motor is a complex process consisting of a series of transitions between different internal motor states. Especially, it is known that backward steps can occur as a consequence of ATP hydrolysis as well as ATP synthesis \cite{Liepelt2007,Liepelt2010}. It is therefore crucial to incorporate detailed kinetic models for the internal conformations into models of molecular traffic, as stressed in several works \cite{Nishinari2005,Brugues2009,Ciandrini2010,Klumpp2008a}. This goal is achieved within model II that is an extension of the ASEP and combines the thermodynamically consistent descriptions of kinesin's stepping presented in \cite{Liepelt2007,Liepelt2010} with the standard exclusion process formalism. To our best knowledge, our model II, as introduced in ref.~\cite{Golubeva2012a}, is the first of its kind to incorporate a kinetic motor model with several mechanochemical cycles into a description of molecular motor traffic.

An obvious quantity to consider when dealing with motors of any kind is their thermodynamic efficiency. For traditional heat engines operating between two thermal baths, the efficiency is constrained by the well-known Carnot's law. Molecular machines, on the other hand, operate in environments at constant temperature, and their efficiency is thus bounded by the thermodynamic limit which is equal to 1. However, both in the case of heat engines and isothermal machines, this lossless limit can only be achieved for reversible processes, or infinitely close to thermodynamic equilibrium. The corresponding power output is therefore zero and thus of limited practical interest. As a consequence, the concept of efficiency at maximum power (EMP) in the context of microscopic engines has recently received considerable attention in the literature since it provides a quantitative measure of the trade-off between power and efficiency \cite{Golubeva2012,Seifert2011a,VandenBroeck2012,Esposito2009}. While these works were concerned with the EMP of single motors, the EMP in systems of interacting molecular motors has only been considered in ref.~\cite{Golubeva2012a}. We exploit model I and II to investigate the effect of mutual motor interactions on the kinetics and thermodynamics in the maximum power regime, and in particular on the EMP. Furthermore, for both models we consider two different types of boundary conditions. The case of open boundary conditions where the system primarily exchanges particles with the reservoir at the ends of the filament appears to be relevant for intracellular transport, where the cell products have to be transported over relatively long distances between different cellular regions. On the other hand, bulk-dominated binding and unbinding, a scenario known as Langmuir kinetics \cite{Klumpp2008a,Parmeggiani2003}, is relevant for experimental \emph{in vitro} studies of molecular motor traffic. Moreover, we consider the case where, besides the mechanical transitions, also the chemical transitions are affected by the ASEP exclusion rule, and discuss the consequences of this, so-called, strong exclusion rule mimicking the possibility that molecular machines can shut down their motor in the presence of high traffic. The idea that non-steric collective effects can suppress dissipation in molecular motor traffic and thereby increase efficiency has been proposed previously in \cite{Shu2004,Slanina2008}. 

We solve the appropriate equations of motion for the various combinations of models, boundary conditions and exclusion types to obtain dynamical quantities such as velocity and hydrolysis rate for fixed chemical input and externally applied mechanical load force. This enables us to characterize the maximal power regime by considering the variation of the power output with the load force for a given value of the chemical driving. Once the optimal load force is known, all the relevant quantities such as EMP or velocity at maximum power can be obtained straightforwardly from the previously calculated dynamics.

We find the remarkable effect that the EMP in many-motor systems is enhanced, as compared to the single motor case, by the mutual exclusion interactions under a wide range of biologically applicable conditions. The enhancement is caused by an altered characteristic force-velocity relation as a consequence of steric motor-motor interactions. Furthermore, in the case of open boundary conditions we observe for both models a trade-off behaviour between efficiency and velocity in the maximum power regime when the force dependence of the unbinding mechanism is altered. This observation has interesting prospects in terms of switching between fast and efficient transport in artificial many-motor systems. In this respect, the so-called 'molecular spiders' are promising examples of highly tunable artificial molecular motors \cite{Lund2010}.

The paper is organized as follows. Sections \ref{sec:modelI} and \ref{sec:maxpowerI} are concerned with model I. In section \ref{sec:modelI} we introduce model I and describe the motor mechanics in the mean-field limit under different boundary conditions, while sec. \ref{sec:maxpowerI} provides a discussion of thermodynamic efficiency, velocity and motor density in the maximum power regime. The following two sections deal with the more detailed model II. In subsections \ref{sec:modelIIa} and \ref{sec:modelIIb} we present the two variations of the model with internal states. We obtain an analytic solution for the open-boundary problem in the mean-field approximation by employing the maximal current principle (MCP) \cite{MCH} as discussed in section \ref{sec:MCP}. In section \ref{sec:maxpowerII} we investigate the EMP of interacting kinesin motors calculated within model II and the corresponding velocity and motor density at maximum power. Finally, section \ref{sec:conclusion} discusses the biological implications of our findings and provides some concluding remarks.

\section{Model I}
\label{sec:modelI}
\begin{figure}
   \centering
   \includegraphics[width=0.8\columnwidth]{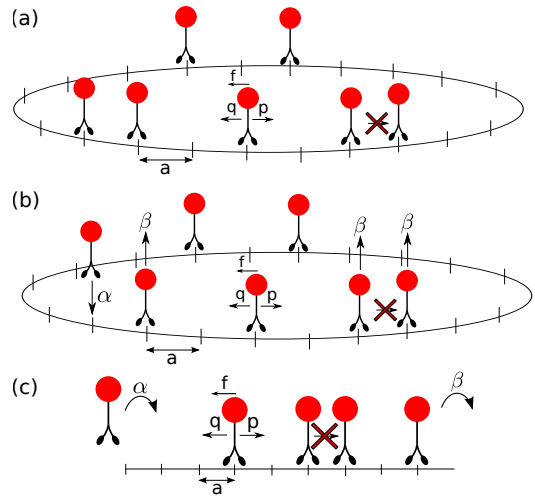}
   \caption{Model I: The standard ASEP model of molecular motor traffic under a load force $f$, where the internal conformational states of the motor are neglected. The motor is thus represented as a particle that can jump forward or backward with rates $p$ and $q$, respectively, if the target side is accessible. The lattice constant corresponds to the motor step size and is denoted by $a$. a) Periodic boundary conditions (PBC). b) PBC and Langmuir kinetics: If a lattice site is empty, a motor can attach to the site with rate $\alpha$. Conversely, a motor bound to the track at a given site can detach from the corresponding site with rate $\beta$. c) Open boundary conditions (OBC): Motors can bind to the filament at the left end with rate $\alpha$, if the first lattice site is empty, and detach from the right end with rate $\beta$.} 
   \label{fig:modelI}
 \end{figure}
We start out by considering a simplified motor traffic model, in which we neglect the internal conformational states of a molecular motor and represent the stepping kinetics as a Poissonian process, see fig. \ref{fig:modelI}. The motor thus moves on the lattice track by performing forward or backward steps of length $a$ with jumping rates $p$ or $q$, respectively. However, the presence of multiple motors on the microtubule gives rise to motor-motor interactions, which we take to be steric in our description. Hence, the motor dynamics is modified by an exclusion rule implying that the stepping can only proceed if the target site in question is unoccupied.

The chemical free energy of the ATP hydrolysis reaction driving the motor can be written in terms of the reactant concentrations as \cite{Howard2001}
\begin{equation}
  \label{eq:dmu}
  \Delta\mu=T\ln \left( \frac{\Keq \cATP}{\cADP \cP} \right),
\end{equation}
where $\Keq$ is the equilibrium constant of the hydrolysis reaction, and $T=\SI{4.1}{\pico\newton \nano\meter}$ is the room temperature (here and in the following, we take $k_B=1$). The released chemical energy is used to perform mechanical work against a constant external load force $f<0$, which represents the effect of the cargo particle on the motor. Since the model motor is tightly coupled, i.e. one ATP molecule is hydrolyzed (synthesized) for every forward (backward) mechanical step, the input work and the output work when completing a forward step are $\win=\Delta\mu$ and $\wout=-f \, a$, respectively. Thermodynamic consistency thus requires that the jumping rates fulfill the local detailed balance condition \cite{Donder1936},
\begin{equation}
  p/q=e^{(\win-\wout)/T}=e^{(\Delta\mu+f a)/T}.
\end{equation}
This relation allows for a parametrization of the rates which for constant ADP and phosphate concentrations reads 
\begin{align}
 \begin{split}
 p&=\omega_0 e^{(\Delta\mu+f a \theta)/T}, \\
 q&=\omega_0 e^{fa(1-\theta)/T},
 \end{split}
\label{eq:rates}
\end{align}
where $\theta$ is the load distribution factor representing the coupling of the mechanical force to the kinetic model parameters.

In the mean-field description the spatial correlations between the different sites are neglected, and the equation of motion governing the bulk dynamics on the lattice reads
\begin{equation}
  \label{eq:modelIbulk}
  \dot{\rho}_j=J_{j+1/2}-J_{j-1/2},
\end{equation}
where $\rho_j$ is the motor density at lattice site $j=1,\dots,N$, and $J_{i+1/2}$ is the mean-field probability current expressed as
 \begin{equation}
   \label{eq:modelIJ}
    J_{j+1/2}=p \rho_j(1-\rho_{j+1})-q \rho_{j+1}(1-\rho_j).
 \end{equation}
We proceed by considering different types of boundary conditions for the system and their effect on the resulting motor dynamics. We focus on periodic boundary conditions (PBC), periodic boundary conditions with Langmuir kinetics (PBC-LK) describing attachment and detachment dynamics in the bulk of the filament, and open boundary conditions (OBC) where motor binding and unbinding occurs at the filament ends. We do not consider the most general, yet technically more involved, case of open boundaries and Langmuir kinetics, since in relevant biological and experimental setups the system can be well approximated by one of the above-mentioned special cases as we argue in sec. \ref{sec:conclusion}.

\subsection{Periodic boundaries (PBC)}
\begin{figure} 
 \psfrag{legendlegend1}[lt][lt][1.]{$f=0$}
 \psfrag{legendlegend2}[lt][lt][1.]{$f=0.5 f_s$}
 \psfrag{legendlegend3}[lt][lt][1.]{$f=0.9 f_s$}
 \psfrag{ylabelvar}[ct][ct][1.]{$\vpbc/a$ (1/s)}
 \psfrag{jpbc}[ct][ct][1.]{$\jpbc$ (1/s)}
 \psfrag{rho}[ct][ct][1.]{$\rho$}
 \psfrag{a}[cB][cB][1.]{{\scriptsize $\textbf{(a)}$}}
 \psfrag{b}[cB][cB][1.]{{\scriptsize $\textbf{(b)}$}}
\centering
\includegraphics[width=.9\columnwidth]{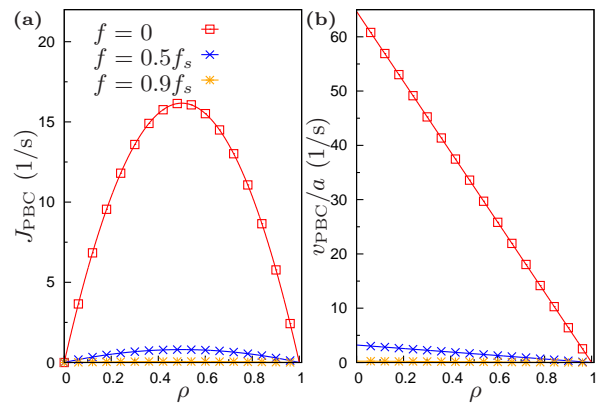}
\caption{Model I (PBC): a) The (mechanical and chemical) current $\jpbc$ as a function of the motor density $\rho$ for $\Delta\mu=20T$ and for three different values of the load force, $f=0,0.5\,f_s,0.9\,f_s$, where $f_s=-\Delta\mu/a$ is the stall force. b) The velocity $\vpbc$ as a function of $\rho$. Legend as in a). Parameter values: $\omega_0=\SI{1.33E-7}{s^{-1}}$, $\theta=0.3$, $a=\SI{8}{\nano\meter}$ (see App. \ref{sec:app1}).}
   \label{fig:pbcI}
 \end{figure}
The simplest description is obtained when neglecting the exchange of molecular motors with a reservoir, i.e. the fact that the motors can attach and detach from the filamenteous track they are moving on. Such a situation corresponds to employing periodic boundary conditions (PBC), see fig. \ref{fig:modelI}a. In this scenario all the lattice sites become equivalent, and the steady-state motor density is thus independent of the lattice site, i.e. $\rho_j\equiv \rho$ for all $j$. The master equation \eqref{eq:modelIbulk} is hence trivially satisfied in the steady-state. 

The probability current takes the form, cf. eq. \eqref{eq:modelIJ}, 
\begin{equation}
  \label{eq:jpbcI}
  \jpbc(\rho)=(p-q)\rho(1-\rho)=v_0 \rho (1-\rho)/a,
\end{equation}
where the density $\rho$ serves as a free parameter. Here, we have introduced the single motor velocity in the absence of interactions and particle exchange with reservoirs, $v_0=a(p-q)$. The motor velocity as a function of $\rho$ is hence given by
\begin{equation}
  \label{eq:vpbcI}
  \vpbc(\rho)=a \jpbc/\rho=a(p-q)(1-\rho)=v_0(1-\rho).
\end{equation}
Fig. \ref{fig:pbcI} illustrates $\jpbc$ and $\vpbc$ for $\Delta\mu=20T$ and for three different values of the load force $f$. In the absence of internal motor states, the molecular traffic model exhibits particle-hole symmetry, i.e. a symmetry upon interchanging $\rho$ and $(1-\rho)$, and $\jpbc$ is thus symmetric around $\rho=1/2$ as can be seen in fig. \ref{fig:pbcI}a.

\subsection{Periodic boundaries and Langmuir kinetics (PBC-LK)}
Next, we consider the effect of binding and unbinding from the filament on the periodic system by introducing Langmuir kinetics (LK), see fig. \ref{fig:modelI}b. In addition to the on-lattice dynamics, a motor can now bind (unbind) from the filament with rate $\alpha$ ($\beta$). The master equation in the steady-state thus assumes the form
\begin{equation}
  \label{eq:modelILK}
  0=\alpha \rho (1-\rho)-\beta\rho.
\end{equation}
 As a result, the motor current remains unchanged, i.e. $\jlkpbc=\jpbc$, while the motor density attains the Langmuir equilibrium value, $\plk=\alpha/(\alpha+\beta)$. We note that the single motor velocity in a non-interacting system with LK is equal to $v_0$, since the interaction with the reservoir only affects the motor run length and not the propagation along the track. The motor thus progresses with the velocity 
 \begin{equation}
   \label{eq:vlkI}
   \vlk=\vpbc(\plk)=v_0(1-\plk)
 \end{equation}
along the filament. Furthermore, we note that the motor velocity is a linear function of the Langmuir density, see also fig. \ref{fig:pbcI}b, which is not the case for the more detailed model IIb as discussed in sec. \ref{sec:modelIIb}. In general, the detachment rate $\beta$ increases with opposing loads \cite{Schnitzer2000,Leidel2012}, and we take the force dependence to be \cite{Liepelt2007,Fisher2001}
 \begin{equation}
   \label{eq:betaf}
   \beta=\beta_0 e^{-\phi fa/T},
 \end{equation}
where $\beta_0$ is the zero-force unbinding rate, and the parameter $\phi$ quantifies the effect of the force on the detachment mechanism. As a result, the motor density is a decreasing function of the applied load, and the motor dynamics in the presence of interactions differs from that of non-interacting motors. 

\subsection{Open boundaries (OBC)}\label{OBC:ss}
\begin{figure} 
 \psfrag{beta}[ct][ct][1.]{$\beta$ $(\si{\second^{-1}})$}
 \psfrag{alpha}[ct][ct][1.]{$\alpha$ $(\si{\second^{-1}})$}
 \psfrag{LD}[ct][ct][1.]{\textcolor{red}{LD}}
 \psfrag{HD}[ct][ct][1.]{\textcolor{red}{HD}}
 \psfrag{MC}[ct][ct][1.]{\textcolor{red}{MC}}
 \psfrag{legendlegend1}[lt][lt][1.]{$f=0$}
 \psfrag{legendlegend2}[lt][lt][1.]{$f=0.5 f_s$}
 \psfrag{legendlegend3}[lt][lt][1.]{$f=0.9 f_s$}
\centering
\includegraphics[width=.9\columnwidth]{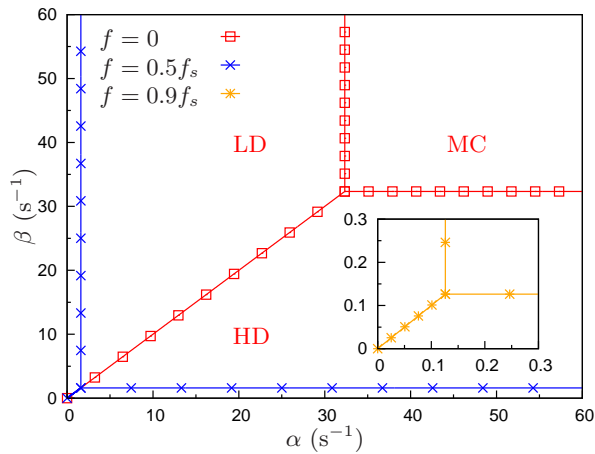}
\caption{Model I (OBC): Phase diagram in the attachment and detachment rates $\alpha$ and $\beta$, respectively, for $\Delta\mu=20T$ and $f=0,0.5\,f_s$ (main figure) and $f=0.9\,f_s$ (inset). Labelling of the phases by LD, HD and MC applies to the $f=0$ phase diagram (squares). Legends and parameters are as in fig. \ref{fig:pbcI}.}
   \label{fig:phasediagI}
 \end{figure}
Finally, we turn our attention to the problem of open boundary conditions (OBC). In this case we assume that the system only exchanges particles with the particle reservoirs at the lattice boundaries as depicted in fig. \ref{fig:modelI}c. Motors can bind to the first lattice site with rate $\alpha$, if the site is unoccupied, and unbind from the last lattice site with rate $\beta$. The master equations for the boundary sites are thus
\begin{align}
\begin{split}
  \dot{\rho}_1&=\alpha (1-\rho_1)-J_{1+1/2} \\
  \dot{\rho}_N&=J_{N-1/2}-\beta \rho_N,
\end{split}
\label{eq:modelIobc}
\end{align}
where $J_{j \pm 1/2}$ is defined in eq.~\eqref{eq:modelIJ}. In the steady-state limit, current conservation entails that the probability current is independent of the lattice site, i.e. $J_{j\pm 1}\equiv J$. In the following we consider the phase diagram for the (mean-field) probability current $J$ in the thermodynamic limit. Since the typical microtubule length, $L=\SI{10}{\micro\meter}$ \cite{Howard2001}, is long compared to the kinesin motor step size $a=\SI{8}{\nano\meter}$, the thermodynamic limit defined by $N=L/a \to \infty$ constitutes a good approximation to the finite-size dynamics. The phase diagram can be obtained by several mean-field techniques such as recursion relations, hydrodynamic equations or the maximal current hypothesis \cite{Derrida1992,Essler1996,Stinchcombe2001,MCH} and happens to coincide with the phase diagram obtained in the thermodynamic limit from the exact analytical solution of the ASEP \cite{ASEP}.

The diagram consists of three regions termed the low-density (LD) phase, the high-density (HD) phase and the maximal current (MC) phase, respectively. In the LD phase the bulk density is determined by the attachment dynamics and equals $\pld=\alpha/(p-q)$. The density in the HD phase is dictated by the detachment dynamics and takes the value $\phd=1-\beta/(p-q)$. Finally, in the MC phase the density is $\pmc=1/2$ and is independent of $\alpha$ and $\beta$. The LD-MC and the HD-MC boundaries are characterized by the critical values $\alpha_c$ and $\beta_c$, respectively, and the particle-hole symmetry implies that $\alpha_c=\beta_c=(p-q)/2$. At the LD-HD boundary, given by $\alpha=\beta$ and $\alpha, \beta \le (p-q)/2$, the low-density and the high-density regions coexist on the lattice and are separated by a fluctuating domain wall. Hence, the location of the phase transitions varies with the chemical input $\Delta\mu$ and the load force $f$, since the hopping rates $p$ and $q$ are functions of $\Delta\mu$ and $f$, cf. eq. \eqref{eq:rates}. In fig. \ref{fig:phasediagI} the phase diagram is depicted for $\Delta\mu=20T$ and for different values of $f$. We note that the MC phase grows with increasing values of the load force. The motor density thus varies with $f$ even when the detachment dynamics is load independent, i.e. $\phi=0$, cf. eq. \eqref{eq:betaf}, as opposed to the system with PBC-LK.
The probability current and the motor velocity are obtained from the phase diagram as $J=\jpbc(\rho_i)$ and $v=\vpbc(\rho_i)$, respectively, where $i$ denotes the appropriate phase, $i=\text{LD}, \text{HD}, \text{MC}$. In summary, using eq. \eqref{eq:vpbcI} for $\vpbc$ the single particle velocity for the interacting system reads
\begin{equation}
\label{eq:vOBCI}
v=
\begin{cases}
v_0-a\alpha & \text{for $\alpha<(p-q)/2$, $\beta>\alpha$ (LD)} \\
a\beta & \text{for $\beta<(p-q)/2$, $\beta<\alpha$ (HD)} \\
v_0/2 & \text{for $\alpha,\beta>(p-q)/2$ (MC)}.
\end{cases}
\end{equation}
It is worth noting that $v$ is always smaller than $v_0$ and attains its maximum in the LD phase as expected, since the jamming is minimal in this phase. Moreover, in the HD phase where it reaches the smallest possible value, the velocity only depends on the detachment rate $\beta$.

\section{Model I: Maximum power regime}
\label{sec:maxpowerI}
The goal of this section is to investigate the operation of our model machine in the maximum power regime when the particle exchange with the reservoirs occurs either at the ends of the filament (OBC) or in the bulk of the system (PBC-LK) as described above. We start out by considering the efficiency at maximum power (EMP) in order to characterize the balance between the output power and thermodynamic efficiency in the presence of motor-motor interactions. Next, we study the corresponding motor velocity and density.   

\subsection{Efficiency at maximum power}
\begin{figure}
 \psfrag{emp2}[ct][ct][1.]{$\etalk^*$}
 \psfrag{emp1}[ct][ct][1.]{$\etaobc^*$}
 \psfrag{Dmu}[ct][ct][1.]{$\Delta\mu/T$}
 \psfrag{a}[cB][cB][1.]{{\scriptsize $\textbf{(a)}$}}
 \psfrag{b}[cB][cB][1.]{{\scriptsize $\textbf{(b)}$}}
 \psfrag{legendlegend1}[lc][lc][1.]{{\scriptsize non-int}}
 \psfrag{legendlegend2}[lc][lc][1.]{{\scriptsize $\phi=0.05$}}
 \psfrag{legendlegend3}[lc][lc][1.]{{\scriptsize $\phi=0.1$}}
 \psfrag{legendlegend4}[lc][lc][1.]{{\scriptsize $\phi=0$}}
\centering
\includegraphics[width=\columnwidth]{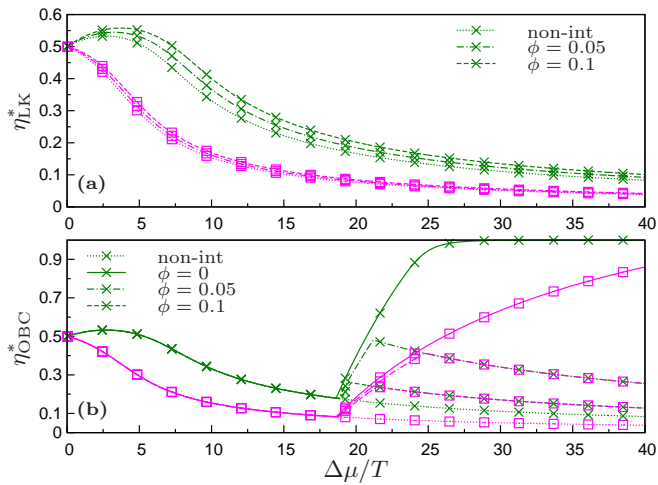}
\caption{Model I: EMP for a single-state model of kinesin interacting through self-exclusion on a periodic lattice with Langmuir kinetics (a), and on an open lattice (b), for two different values of the load distribution factor, $\theta=0.3$ (green, crosses) and $\theta=0.65$ (magenta, squares), and different load dependencies of the detachment rate, see eq. \eqref{eq:betaf}, as indicated in the legends. For comparison, the EMP for non-interacting motors with the same parameter values is shown with dotted lines. Parameter values: $\omega_0=\SI{1.33E-7}{s^{-1}}$, $a=\SI{8}{\nano\meter}$, $\alpha=\SI{5}{\second^{-1}}$, $\beta_0=\SI{3}{\second^{-1}}$ (see App. \ref{sec:app1}). The values for $\theta$ are taken from \cite{Liepelt2007} and correspond to two independent experiments on the kinesin motor.}
   \label{fig:modelIemp}
 \end{figure}
Since the chemical and mechanical currents are tightly coupled in model I, the hydrolysis rate $r$ describing the number of consumed ATP molecules per time unit is proportional to the motor velocity, i.e. $r=v/a$. The input power $\Pin=\Delta\mu r$ is thus proportional to the delivered output power $\Pout=-fv$, and the efficiency $\eta$ of the motor is independent of the mechanical and chemical currents,
\begin{equation}
  \label{eq:etaI}
  \eta=\frac{\Pout}{\Pin}=\frac{-fv}{\Delta\mu r}=\frac{-fa}{\Delta\mu}=\frac{f}{f_s}.
\end{equation}
Here, $f_s=-\Delta\mu/a$ is the stalling force of the motor for which the forward and backward rates are equal, $p=q$, and the velocity vanishes. It is clear from eq. \eqref{eq:etaI} that the maximum efficiency, $\eta=1$, is obtained under stalling conditions corresponding to equilibrium where the power output vanishes. A more relevant quantity to consider is therefore the efficiency at maximum power as discussed in the Introduction. For fixed $\Delta\mu$ we thus calculate the motor efficiency when the power is optimized with respect to the load force. The equation for the maximizing force $f^*$ thus reads
\begin{equation*}
  \left. \der{\Pout}{f} \right|_{f^*} =\left. \der{}{f}(-fv) \right|_{f^*}= \left. -\left( v+f \der{v}{f}\right) \right|_{f^*}=0,
\end{equation*}
and the EMP is simply obtained as $\eta^*=f^*/f_s$. The above procedure is then repeated for increasing values of the ATP concentration, and hence $\Delta\mu$. 
It is worth noting that the concept of EMP is different from the maximal possible efficiency at fixed $\Delta \mu$, as given by the condition $ \partial\eta/\partial f=0$ and studied in, e.g., ref.~\cite{Liepelt2010}.
The results obtained for the EMP for model I are reported in fig. \ref{fig:modelIemp} for different values of the load sharing parameter $\theta$ and for different load dependencies of the detachment rate characterized by the parameter $\phi$. 

\begin{figure}
 \psfrag{Dmu1}[ct][ct][1.]{{\scriptsize $\Delta\mu=10T$}}
 \psfrag{Dmu2}[ct][ct][1.]{{\scriptsize $\Delta\mu=20T$}}
 \psfrag{pout}[ct][ct][1.]{$\Pout$ (pN nm/s)}
 \psfrag{f}[ct][ct][1.]{$f$ (pN)}
 \psfrag{Ia}[cB][cB][1.]{{\scriptsize $\textbf{(a)}$}}
 \psfrag{Ib}[cB][cB][1.]{{\scriptsize $\textbf{(b)}$}}
\centering
\includegraphics[width=\columnwidth]{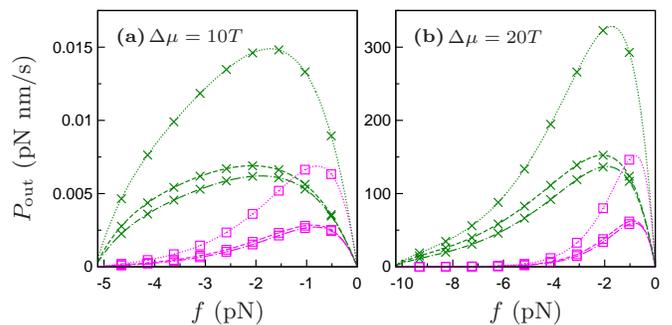}
\caption{Model I (PBC-LK): the output power $\Pout$ as a function of the external load force $f$ for two different values of the chemical input, $\Delta\mu=10T$ (a) and $\Delta\mu=20T$ (b), and for different values of the load dependence parameters $\theta$ and $\phi$. Legends and parameters are as in fig. \ref{fig:modelIemp}.}
   \label{fig:modelILK_pout}
 \end{figure}
The EMP, $\etalk^*$, for the system with PBC-LK is shown in fig. \ref{fig:modelIemp}a together with the EMP for the corresponding non-interacting system, $\eta_0^*$. We observe an increase in $\etalk^*$ with respect to $\eta_0^*$ due to a change in the characteristic force-velocity relation $v(f)$, and hence $\Pout(f)$, as illustrated in fig. \ref{fig:modelILK_pout}. Since both $v_0$ and $\plk$ are decreasing functions of the load, it follows from eq. \eqref{eq:vlkI} that the velocity decreases more slowly with the force in the presence of steric interactions. This, in turn, leads to a higher optimal force $f^*$, as compared to the non-interacting system, and hence to a higher EMP. The enhancement of the EMP occurs for all values of $\theta$ and all non-zero values of $\phi$. For $\phi=0$ we have $\vlk \propto v_0$ because $\plk$ is constant, and $\etalk^*$ is thus equal to $\eta_0^*$. Hence, larger values of $\phi$ result in a greater effect of motor-motor interactions on the EMP. Furthermore, the increase in the EMP is more pronounced for small values of $\theta$, as has been discussed previously in \cite{Golubeva2012a}. In the linear regime given by $\Delta\mu/T\to 0$ we recover the well-known result $\eta^*\to 1/2$ for tightly coupled systems \cite{Golubeva2012,Seifert2011a,VandenBroeck2012}. Moreover, $\etalk^* \to \eta_0^*$ for $\Delta\mu/T \to \infty$, since the single-motor velocity $v_0$ is the dominating contribution to $\vlk$, eq. \eqref{eq:vlkI}, for large $\Delta\mu$. 

\begin{figure}
 \psfrag{MC}[ct][ct][1.]{{\scriptsize MC}}
 \psfrag{HD}[ct][ct][1.]{{\scriptsize HD}}
 \psfrag{LD}[ct][ct][1.]{{\scriptsize LD}}
 \psfrag{Dmu1}[ct][ct][1.]{{\scriptsize $\Delta\mu=10T$}}
 \psfrag{Dmu2}[ct][ct][1.]{{\scriptsize $\Delta\mu=21T$}}
 \psfrag{Dmu3}[ct][ct][1.]{{\scriptsize $\Delta\mu=26T$}}
 \psfrag{phi1}[cB][cB][1.]{{\scriptsize non-int}}
 \psfrag{phi2}[cB][cB][1.]{{\scriptsize $\phi=0$}}
 \psfrag{phi3}[cB][cB][1.]{{\scriptsize $\phi=0.05$}}
 \psfrag{pout}[ct][ct][1.]{$\Pout$ (pN nm/s)}
 \psfrag{f}[ct][ct][1.]{$f$ (pN)}
 \psfrag{Ia}[cB][cB][1.]{{\scriptsize $\textbf{(Ia)}$}}
 \psfrag{Ib}[cB][cB][1.]{{\scriptsize $\textbf{(Ib)}$}}
 \psfrag{Ic}[cB][cB][1.]{{\scriptsize $\textbf{(Ic)}$}}
 \psfrag{IIa}[cB][cB][1.]{{\scriptsize $\textbf{(IIa)}$}}
 \psfrag{IIb}[cB][cB][1.]{{\scriptsize $\textbf{(IIb)}$}}
 \psfrag{IIc}[cB][cB][1.]{{\scriptsize $\textbf{(IIc)}$}}
 \psfrag{IIIa}[cB][cB][1.]{{\scriptsize $\textbf{(IIIa)}$}}
 \psfrag{IIIb}[cB][cB][1.]{{\scriptsize $\textbf{(IIIb)}$}}
 \psfrag{IIIc}[cB][cB][1.]{{\scriptsize $\textbf{(IIIc)}$}}
\centering
\includegraphics[width=\columnwidth]{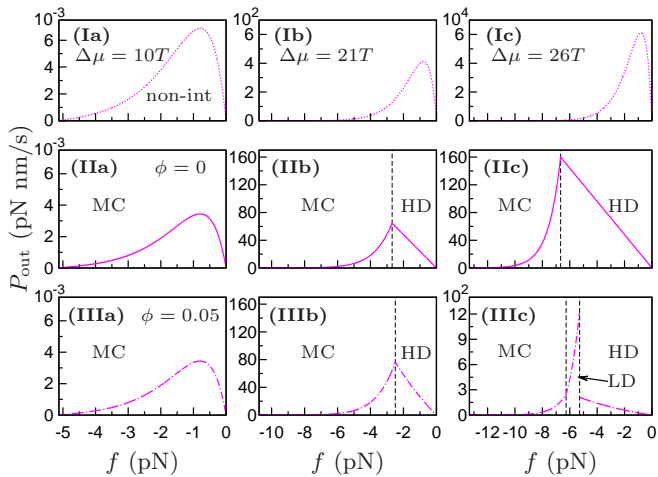}
\caption{Model I (OBC): The output power $\Pout$ as a function of the external load force $f$ for three different values of the chemical input, $\Delta\mu=10T$ (a),$\Delta\mu=21T$ (b) and $\Delta\mu=26T$ (c). The  curves for the non-interacting system are shown in the figures in row (I). The other figures correspond to the interacting system with different values of the load dependence parameter for detachment, $\phi=0$ (II) and $\phi=0.05$ (III).  Here, we have taken $\theta=0.65$, while the rest of the parameters and the legends are as in fig. \ref{fig:modelIemp}. Note the different scales on the y-axes.}
   \label{fig:modelIOBC_pout}
 \end{figure}
Fig. \ref{fig:modelIemp}b shows the EMP, $\etaobc^*$, for the system interacting with the environment at the boundaries together with the corresponding EMP in the absence of interactions, $\eta_0^*$. The EMP exhibits different behaviour with $\Delta\mu$ depending on the value of $\phi$. If the detachment process depends weakly on the load force (or in the limit $\phi=0$), the EMP displays two different regimes. When the chemical free energy is smaller than the critical value $\Delta\mu_{c,1}$ (see App. \ref{sec:app2} for a discussion of the critical values), the maximum power is reached in the MC phase. Since the velocity in this phase is proportional to $v_0$, cf. eq.~\eqref{eq:vOBCI}, we obtain $\etaobc^*=\eta_0^*$ in this regime, see fig. \ref{fig:modelIOBC_pout}I-IIa. For $\Delta\mu>\Delta\mu_{c,1}$ the maximum of $\Pout$ is no longer achieved in the MC phase, but rather at the dynamical phase transition between the HD and the MC phase, as depicted in figs. \ref{fig:modelIOBC_pout}I-IIb. As a consequence, the optimal force $\fobc^*$ is larger than the corresponding force $f_0^*$ for the non-interacting system, which in turn leads to a higher EMP, in analogy to the observation made for the system with PBC-LK. In particular, for $\Delta\mu/T\to\infty$, the HD-MC phase transition occurs infinitely close to stall, which implies that $\etaobc^*\to 1$ in this limit. We note, however, that this limit is caused by the symmetry of the phase diagram and is unphysical. We shall see in the following that under more realistic conditions, such as dissipation by futile hydrolysis, as considered in model II, or non-negligible load dependence of $\beta$, the EMP decays to zero as $\Delta\mu$ approaches infinity.

Besides the two regimes characterized above, the EMP exhibits a third, qualitatively different, regime whenever $\phi \neq 0$, see figs. \ref{fig:modelIOBC_pout}IIIa-c. When $\Delta\mu$ is larger than the critical value $\Delta\mu_{c,2}$ with $\Delta\mu_{c,2}>\Delta\mu_{c,1}$ (see App. \ref{sec:app2} for a discussion of the critical value $\Delta\mu_{c,2}$), the motors operate at maximum power at the discontinuous phase transition between the HD and the LD phases as shown in fig. \ref{fig:modelIOBC_pout}IIIc. Hence, the maximizing force only depends on $\phi$,
\begin{equation}
 \fobc^*=-\log(\alpha\beta_0)T/\phi a,
 \label{eq:fOBCs}
\end{equation}
and is (numerically) larger than $f_0^*$. Since the stalling force increases with $\Delta\mu$ and $\fobc^*$ is constant, the EMP thus goes to zero for $\Delta\mu \to \infty$ as illustrated in fig. \ref{fig:modelIemp}b. Furthermore, the EMP decreases with increasing $\phi$. It is worth noting that the dependence on $\phi$ is contrary to the one obtained for PBC+LK, where the EMP, and hence the enhancement in the EMP, increased when increasing $\phi$. Moreover, it follows from eq. \eqref{eq:fOBCs} that the EMP $\etaobc^*$ in this regime is independent of $\theta$, while the enhancement in the EMP represented by the ratio $\etaobc^*/\eta_0^*$ increases with increasing $\theta$, which is again at variance with the behaviour observed for PBC+LK. Thus, the detachment dynamics at the boundaries and in the bulk, respectively, affects the thermodynamics of molecular machines differently. 

In conclusion, for $\Delta\mu > \Delta\mu_{c,1}$, we observe, as for the case of PBC-LK, that motor-motor interactions lead to an increase in the EMP, as compared to the non-interacting system, due to a change in the characteristic response of the velocity to the external driving.

\subsection{Velocity and density at maximum power}
\begin{figure}
 \psfrag{vstar1}[ct][ct][1.]{$\vlk^*$ (nm/s)}
 \psfrag{vstar2}[ct][ct][1.]{$\vobc^*$ (nm/s)}
 \psfrag{Dmu}[ct][ct][1.]{$\Delta\mu/T$}
 \psfrag{a}[cB][cB][1.]{{\scriptsize $\textbf{(a)}$}}
 \psfrag{b}[cB][cB][1.]{{\scriptsize $\textbf{(b)}$}}
 \psfrag{legendlegend1}[lc][lc][1.]{{\scriptsize non-int}}
 \psfrag{legendlegend2}[lc][lc][1.]{{\scriptsize $\phi=0.05$}}
 \psfrag{legendlegend3}[lc][lc][1.]{{\scriptsize $\phi=0.1$}}
 \psfrag{legendlegend4}[lc][lc][1.]{{\scriptsize $\phi=0$}}
\centering
\includegraphics[width=\columnwidth]{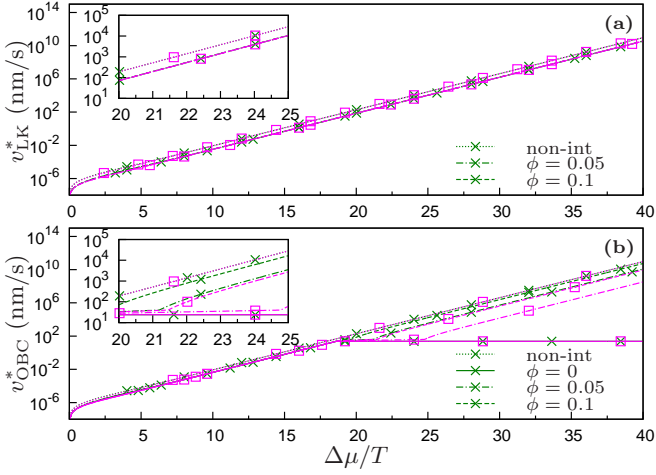}
\caption{Model I: The velocity at maximum power corresponding to the EMP shown in fig.~\ref{fig:modelIemp}. Legends and parameters are as in fig. \ref{fig:modelIemp}. The insets are enlargements of the region $\Delta\mu/T=20-25$ of the corresponding figure.}
   \label{fig:modelIvstar}
 \end{figure}
\begin{figure}
 \psfrag{rhostar1}[ct][ct][1.]{$\plk^*$}
 \psfrag{rhostar2}[ct][ct][1.]{$\pobc^*$}
 \psfrag{Dmu}[ct][ct][1.]{$\Delta\mu/T$}
 \psfrag{a}[cB][cB][1.]{{\scriptsize $\textbf{(a)}$}}
 \psfrag{b}[cB][cB][1.]{{\scriptsize $\textbf{(b)}$}}
 \psfrag{legendlegend1}[lc][lc][1.]{{\scriptsize non-int}}
 \psfrag{legendlegend2}[lc][lc][1.]{{\scriptsize $\phi=0.05$}}
 \psfrag{legendlegend3}[lc][lc][1.]{{\scriptsize $\phi=0.1$}}
 \psfrag{legendlegend4}[lc][lc][1.]{{\scriptsize $\phi=0$}}
\centering
\includegraphics[width=\columnwidth]{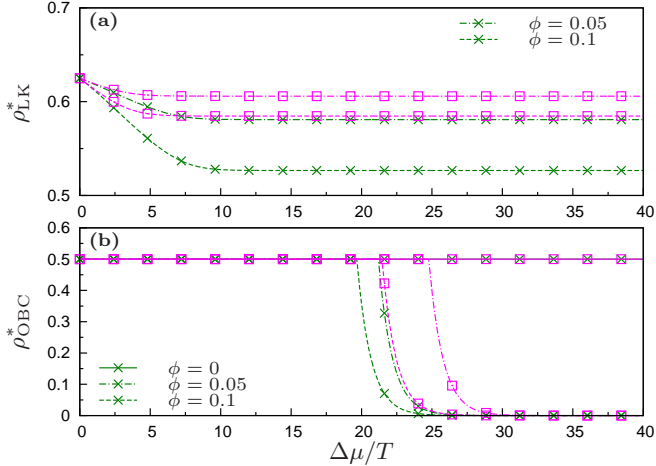}
\caption{Model I: The density at maximum power corresponding to the EMP shown in fig. \ref{fig:modelIemp}. Legends and parameters are as in fig. \ref{fig:modelIemp}. Note the different scales on the y-axes.}
   \label{fig:modelIrhostar}
 \end{figure}

We proceed by discussing the velocity at maximum power (VMP) and the density at maximum power (DMP) depicted in figs. \ref{fig:modelIvstar} and \ref{fig:modelIrhostar}, respectively, for the two cases of PBC-LK and OBC. When the particle exchange occurs in the bulk, fig. \ref{fig:modelIvstar}a shows that the VMP, $\vlk^*$, is smaller than the corresponding VMP for the non-interacting case, $v_0^*=v_0(f_0^*)$ by approximately a factor of three for most values of $\Delta\mu$. Such relatively weak effect of the mutual interactions on the VMP is caused by the essentially constant DMP, see fig. \ref{fig:modelIrhostar}a. Since the optimal force $f^*$ is roughly constant with $\Delta\mu$, see fig. \ref{fig:modelILK_pout}, the DMP $\plk^*=\alpha/(\alpha+\beta(f^*))$ only varies weakly with $\Delta\mu$. Furthermore, since $f^*$ is close to the corresponding force in the absence of interactions, $f_0^*$, the resulting VMP is smaller than $v_0^*$ by approximately a factor of $1-\plk^*$, see eq.~\eqref{eq:vlkI}, and is hence roughly proportional to $v_0^*$. Moreover, we note that the DMP is rather insensitive to changes in the load dependence parameters $\theta$ and $\phi$ when these parameters attain realistic values, see fig. \ref{fig:modelIrhostar}a. This in turn leads to a negligible dependence of the VMP on $\theta$ and $\phi$.      

The VMP in the case of OBC is depicted in fig. \ref{fig:modelIvstar}b and exhibits three different regimes in analogy to the EMP, see fig. \ref{fig:modelIemp}b. In the first regime, characterized by $\Delta\mu<\Delta\mu_{c,1}$, maximum power is achieved in the MC phase for $f^*=f_0^*$, and the VMP is thus simply $\vobc^*=v_0^*/2$, cf. eq. \eqref{eq:vOBCI}. The corresponding DMP is $\pobc^*=1/2$ as shown in fig. \ref{fig:modelIrhostar}b. For $\Delta\mu_{c,1}<\Delta\mu<\Delta\mu_{c,2}$ (or $\Delta\mu_{c,1}<\Delta\mu$ for $\phi=0$), the system operates at maximum power at the MC-HD boundary. The DMP is thus again $\pobc^*=1/2$, while the VMP is given by $\vobc^*=a\beta(f^*)$ and is considerably smaller than the velocity $v_0^*$ of non-interacting motors in the maximum power regime. In the third regime defined by $\Delta\mu_{c,2}<\Delta\mu$, the power output is optimal at the LD-HD phase transition. Here, the output power and the motor velocity as functions of $f$ are discontinuous due to a discontinuity in the density. However, keeping in mind that the maximum power regime corresponds to a specific value of the force, and a molecular motor might operate at a slightly different force, one may relax the requirement down to, say, 90$\%$ of maximum power. This places the system in the low density regime, and  in figs. \ref{fig:modelIvstar}b and \ref{fig:modelIrhostar}b we thus report for $\Delta\mu_{c,2}<\Delta\mu$ the low-density values for the VMP, $\vobc^*=\vld(f\to f^*)$, and DMP, $\pobc^*=\pld(f\to f^*)$, respectively. The transition to this regime is accompanied by a drastic decrease in the DMP. It is also worth noting that the VMP in this region increases with decreasing values of $\theta$. Furthermore, it is interesting to observe that the EMP, as well as the enhancement in the EMP, increases significantly when decreasing $\phi$, while the VMP decreases notably when $\phi$ is decreased. Thus, in artificial many-motor systems, altering the force dependence of the detachment dynamics can serve as a control mechanism for the trade-off between efficient and fast transport.   

\section{Model II}
\label{sec:modelII}
We will now generalize model I to include several biochemical motor states. In order to achieve this goal we consider two different network models for kinesin's mechanochemical cycles and incorporate them into the standard ASEP. Model IIa is a six-state model introduced in \cite{Liepelt2007} to describe a single kinesin motor operating under a constant external load force, while  model IIb represents  an extension of the previous model \cite{Liepelt2007} to seven states \cite{Liepelt2010}. As shown in \cite{Liepelt2010}, the two single motor models provide similar predictions for the thermodynamics of kinesin for $\Delta\mu \gtrsim 20T$, which includes the biologically relevant range for the free energy, $\Delta\mu/T \sim 20-25$ \cite{Hackney2005}. Since model IIa, due to the presence of only one mechanical stepping transition, admits an analytic solution for OBC by using the maximal current principle, we will primarily use model IIa to investigate the effect of interactions on the open-boundary dynamics. When the Langmuir kinetics dominates, however, a detailed description of the detachment dynamics is necessary, which requires the presence of the seventh state, and model IIb is therefore the appropriate to consider in this case. Furthermore, for both model IIa and IIb we explore the possibility that steric interactions affect the internal conformational motor dynamics leading to mechanical stepping, as well as the stepping transitions themselves. In the following we refer to this type of exclusion rule as strong exclusion.   

Below we describe the two models IIa and IIb in detail and discuss the mean-field solutions of the models.

\subsection{Model IIa}
\label{sec:modelIIa}
\begin{figure}
   \centering
   \includegraphics[width=\columnwidth]{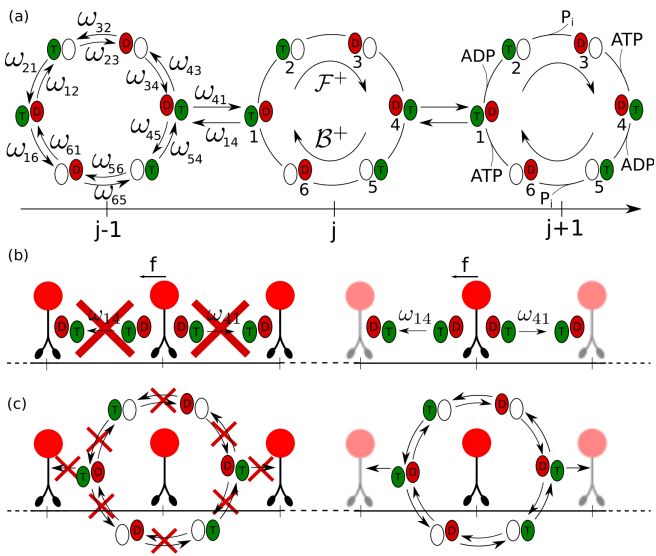}
   \caption{Model IIa: a)  The six-state network model of kinesin's mechanochemical cycles that is introduced in \cite{Liepelt2007} and is used to model the internal dynamics of kinesin motors under the assumption of one mechanical stepping transition. Each state is characterized by the individual states of the two motor heads that can either be empty, contain bound ATP (T) or bound ADP (D). The system performs transitions from state $i$ to state $j$ with rate $\omega_{ij}$. The forward and backward stepping dicycles, $\mathcal{F}^+$ and $\mathcal{B}^+$, respectively, are indicated in the figure (see text). b) Mechanical exclusion (ME): When incorporating the mechanochemical cycles depicted in a) into the standard ASEP (see fig. \ref{fig:modelI}) the exclusion rule affects the mechanical stepping transitions between states 1 and 4. c) Strong exclusion (SE): In this case the internal chemical transitions as well as the mechanical transitions are prohibited in the presence of neighbouring motors.} 
   \label{fig:modelIIa}
 \end{figure}
This model is based on the six-state network of biochemical states introduced in \cite{Liepelt2007} and depicted in fig. \ref{fig:modelIIa}a. Kinesin is a two-headed motor, and each of the two motor heads goes through an ATP hydrolysis cycle. Since the rate of phosphate release from a motor head is limited by the preceding hydrolysis reaction, i.e. $\phos$ is released immediately after ATP hydrolysis, these two subsequent transitions can be combined into a single one. Each head can therefore either be empty, contain bound ATP or bound ADP. The discrete state space of kinesin is thus composed of $3^2=9$ states that differ in chemical composition. However, in Ref. \cite{Liepelt2007} Liepelt and Lipowsky argue that the chemical processes in the heads are coordinated, and hence that only states where the two heads have different chemical composition should be connected in the network. They thus arrive at the reduced state space consisting of six chemical states. The motor hydrolyzes an ATP molecule and performs a forward step through the $\mathcal{F}^+=|12341 \rangle$ dicycle, while the  $\mathcal{B}^+=|45614 \rangle $ dicycle represents ATP hydrolysis leading to backward stepping, see fig. \ref{fig:modelIIa}a. Furthermore, the network includes the futile hydrolysis dicycle  $\mathcal{D}^+=| 1234561 \rangle$, where two ATP molecules are consumed while no stepping, and hence output work, is performed by the motor. 
In order to satisfy the constraints dictated by the local detailed balance requirement, the network model must contain the reverse dicycles $\mathcal{F}^-=|14321 \rangle$,  $\mathcal{B}^-=|41654 \rangle $ and $\mathcal{D}^-=| 1654321 \rangle$, in which ATP synthesis takes place.

The mechanical steps where the two heads switch positions on the track correspond to the transitions between states 1 and 4 as shown in fig. \ref{fig:modelIIa}. The standard ASEP exclusion rule modifies these mechanical stepping transitions, see fig. \ref{fig:modelIIa}b, while all the other transitions representing internal conformational changes remain unaffected. Furthermore, as for the standard ASEP we allow the motors to bind to the microtubule at the left end with rate $\alpha$ and detach at the right end with rate $\beta$. In the mean-field approximation the evolution of the system in the bulk is described by the master equations,
\begin{align}
\begin{split}
  \dot{\rho}_j^1=&\omega_{41}\rho_{j-1}^4(1-\rho_j)-\omega_{14}\rho_j^1 (1-\rho_{j-1}) \\
                &+ \omega_{21}\rho_j^2-\omega_{12}\rho_j^1+\omega_{61}\rho_j^6-\omega_{16}\rho_j^1 \\ 
  \dot{\rho}_j^4=& -\omega_{41}\rho_j^4(1-\rho_{j+1})+\omega_{14}\rho_{j+1}^1(1-\rho_j) \\
                 & -\omega_{43}\rho_j^4+\omega_{34}\rho_j^3+\omega_{54}\rho_j^5-\omega_{45}\rho_j^4    \\
  \dot{\rho}_j^k=& - \omega_{k\, k-1}\rho_j^k+\omega_{k-1 \, k}\rho_j^{k-1} \\
                & +\omega_{k \, k+1}\rho_j^k-\omega_{k+1 \, k}\rho_j^{k+1} \quad \text{for } k\neq 1,4,  
\end{split}
\label{eq:MF}
\end{align}
with the obvious notation $k+1=1$ if $k=6$ and $k-1=6$ if $k=1$. Here, $\omega_{ij}$ is the transition rate for going from state $i$ to state $j$. The density of motors in the chemical state $i$ at lattice site $j$ is denoted by $\rho_j^i$, while $\rho_j=\sum_{i=1}^6 \rho_j^i$ is the total motor density at site $j$ \footnote{Strictly speaking, $\rho_j^i$ is an occupation probability, while $\rho_j^i/a$ is the density. However, we will for simplicity refer to $\rho_j^i$ as density.}. Note that the equations for $\rho_j^1$ and $\rho_j^4$ are modified as compared to the non-interacting system due to mutual exclusion. 
Open boundary conditions entail that eqs.~\eqref{eq:MF} should be modified for the first and the last lattice sites, analogously to the case of model I discussed in section~\ref{OBC:ss}. However, we will show in the following that only the total incoming (outgoing) probability current at the left (right) boundary enters the mean-field solution. Hence, the detailed dynamics of the motors at the boundary sites is unimportant as long as the flux of motors entering or leaving the microtubule remains unchanged. 

We solve the model introduced above under steady-state conditions and in the thermodynamic limit using a mean-field approach known as the maximum current principle (MCP) \cite{MCH}. According to the MCP, the boundaries of the lattice are substituted by reservoirs of particles, and the dynamics between the reservoirs and the lattice is assumed to be identical to that in the bulk. The (constant) densities of the left and the right reservoirs are denoted by $\rho_l$ and $\rho_r$, respectively. The mechanical probability current through the open system in the thermodynamic limit, $J^{\text{m}}$, is then predicted by the MCP to be \cite{MCH}
\begin{equation}
  \label{eq:mch}
  J^{\text{m}}=
\begin{cases}
\max \limits_{\rho\in [\rho_r,\rho_l]} \jmpbc(\rho) & \text{for $\rho_l>\rho_r$} \\
\min \limits_{\rho\in [\rho_l,\rho_r]} \jmpbc(\rho) & \text{for $\rho_l<\rho_r$}, 
\end{cases}
\end{equation}
where $\jmpbc(\rho)$ is the steady-state mechanical probability current through a corresponding homogeneous system with periodic boundary conditions (PBC) and density $\rho$. Furthermore, the MCP implicitly states that the bulk density of the system takes on the value that maximizes (or minimizes) the mechanical current through the lattice as specified by eq. \eqref{eq:mch}. Hence, we proceed by considering the problem of motor traffic on a periodic lattice.   

\subsubsection{Periodic boundaries}
For the periodic system all the lattice sites are equivalent, and the densities thus become independent of the position on the track, $\rho_j^i \equiv \rho^i$. In this case, the governing equations of motion, eq. \eqref{eq:MF}, admit an analytical solution for $\rho^i$ in the steady-state limit, i.e. for $\dot{\rho}^i=0$. Since the master equations are linearly dependent, the homogeneous density $\rho=\sum_{i=1}^6 \rho^i$ serves as an independent parameter under PBC. Each mechanochemical cycle $\mathcal{C}$ then carries a unique probability current,
\begin{equation}
 J(\mathcal{C})=J(\mathcal{C^+})-J(\mathcal{C^-}), 
\end{equation}
where $\mathcal{C}$ is either the forward stepping cycle $\mathcal{F}$, the backward stepping cycle $\mathcal{B}$ or the futile hydrolysis cycle $\mathcal{D}$. The mechanical probability current on the lattice can hence be expressed as
\begin{figure} 
 \psfrag{legendlegend1}[lt][lt][1.]{$f=0$}
 \psfrag{legendlegend2}[lt][lt][1.]{$f=0.5 f_s$}
 \psfrag{legendlegend3}[lt][lt][1.]{$f=0.9 f_s$}
 \psfrag{ylabelvar}[ct][ct][1.]{$\vpbc/a$, $\, \rpbc$ (1/s)}
 \psfrag{var}[ct][ct][1.]{$\vpbc/a \, \rpbc$}
 \psfrag{jpbc}[ct][ct][1.]{$\jmpbc$, $\jcpbc$ (1/s)}
 \psfrag{rho}[ct][ct][1.]{$\rho$}
 \psfrag{a}[cB][cB][1.]{{\scriptsize $\textbf{(a)}$}}
 \psfrag{b}[cB][cB][1.]{{\scriptsize $\textbf{(b)}$}}
\centering
\includegraphics[width=\columnwidth]{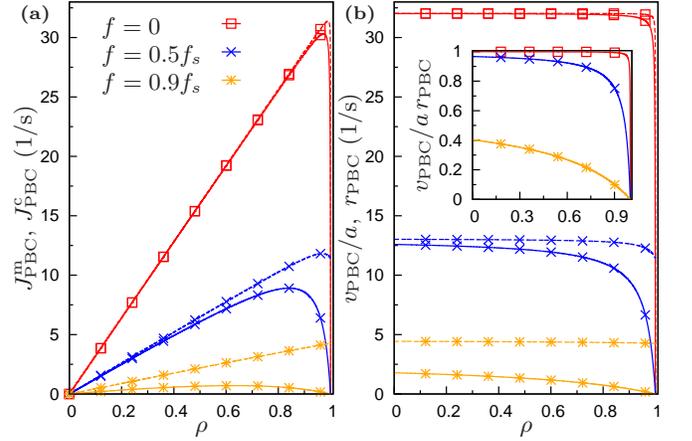}
\caption{Model IIa (PBC): a) The mechanical current $\jmpbc$ (solid lines) and the chemical current $\jcpbc$ (dashed lines) as a function of the motor density $\rho$ for $\Delta\mu=20T$ and for three different values of the load force, $f=0,0.5\,f_s,0.9\,f_s$, where $f_s$ is the stall force. b) The velocity $\vpbc$ and the hydrolysis rate $\rpbc$ as a function of $\rho$. Legend as in a). Inset: The ratio $\vpbc/a \, \rpbc$ that gives the average number of steps that the motor moves per hydrolyzed ATP molecule. This ratio also represents the coupling degree of the motor, where 1 indicates tight coupling. The parameter values used are those obtained in \cite{Liepelt2007} for the Carter and Cross experiment \cite{Carter2005}. }
   \label{fig:pbc}
 \end{figure}
\begin{align}
\begin{split}
 \jmpbc(\rho)&=J(\mathcal{F})-J(\mathcal{B}) \\
             &= (\omega_{34}\rho^3-\omega_{43}\rho^4)-(\omega_{45}\rho^4-\omega_{54}\rho^5),
\end{split}
\label{eq:jmpbc}
\end{align}  
while the chemical current accounting for fuel consumption becomes
\begin{align}
\begin{split}
 \jcpbc(\rho)&=J(\mathcal{F})+J(\mathcal{B}) \\
             &= (\omega_{34}\rho^3-\omega_{43}\rho^4)+(\omega_{45}\rho^4-\omega_{54}\rho^5),
\end{split}
\label{eq:jcpbc}
\end{align}
where the second equalities follow from one specific representation of $J(\mathcal{F})$ and $J(\mathcal{B})$, which is, however, not unique due to current conservation. Figure \ref{fig:pbc}a illustrates $\jmcpbc(\rho)$ for $\Delta\mu=20T$ and for different values of the load force. We note that $\jmpbc$ is not symmetric around the value $\rho=1/2$, as it is in the case of the standard ASEP \cite{ASEP}, since the particle-hole symmetry is broken by the presence of internal conformational states, as also pointed out in \cite{Ciandrini2010,Klumpp2008a}.

The motor velocity, $\vpbc$, and the ATP hydrolysis rate, $\rpbc$, can be obtained from the probability currents as
\begin{align}
\begin{split}
  \vpbc(\rho)&=a\jmpbc(\rho)/\rho \\
  \rpbc(\rho)&=\jcpbc(\rho)/\rho
\end{split}
\label{eq:vpbc_rpbc}
\end{align}
and are plotted in fig. \ref{fig:pbc}b. Equations \eqref{eq:jmpbc}--\eqref{eq:vpbc_rpbc} underline that kinesin is a loosely coupled motor since the average number of steps per unit time, $\vpbc/a$, is in general smaller than the hydrolysis rate $\rpbc$ as depicted in the inset of fig. \ref{fig:pbc}b. The motor velocity vanishes, i.e. $\vpbc=0$, under stalling conditions characterized by $J(\mathcal{F})=J(\mathcal{B})$, see eq. \eqref{eq:jmpbc}. For fixed chemical input $\Delta\mu$ this condition determines the stall force $f_s$, which is identical to the one obtained for non-interacting motors. Futile hydrolysis implies, however, that $r$ is non-vanishing at stall, and the coupling between the chemistry and the mechanical motion is low in this limit, see inset of fig. \ref{fig:pbc}b.

\subsubsection{Open boundaries: Maximal current principle}
\label{sec:MCP}
\begin{figure} 
 \psfrag{ps}[cB][cB][1.]{{\footnotesize $\bar{\rho}$}}
 \psfrag{pl}[cB][cB][1.]{{\footnotesize $\rho_l$}}
 \psfrag{pr}[lB][lB][1.]{{\footnotesize $\rho_r$}}
 \psfrag{prm}[lB][lB][1.]{\textcolor{blue}{{\footnotesize $\prmax$}}}
 \psfrag{plm}[cB][cB][1.]{\textcolor{blue}{{\footnotesize $\plmin$}}} 
 \psfrag{jpbc}[ct][ct][1.]{$\jmpbc(\rho)$ (1/s)}
 \psfrag{rho}[ct][ct][1.]{$\rho$}
 \psfrag{a}[lB][lB][1.]{{\scriptsize $\textbf{(a)}$ } LD}
 \psfrag{b}[lB][lB][1.]{{\scriptsize $\textbf{(b)}$ } HD}
 \psfrag{c}[lB][lB][1.]{{\scriptsize $\textbf{(c)}$ } MC}
\centering
\includegraphics[width=\columnwidth]{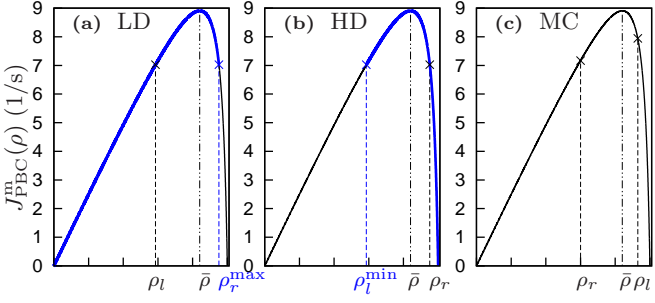}
\caption{OBC: Illustration of the MCP prediction, eq. \eqref{eq:mch}. a) Low-density (LD) phase with bulk density $\pld=\rho_l$, $0<\rho_l<\bar{\rho}$ and $0<\rho_r<\prmax$ (see text). The interval for $\rho_r$ is indicated by a wide (blue) line in the figure. b) High-density (HD) phase with bulk density $\phd=\rho_r$, $\bar{\rho}<\rho_r<1$ and $\plmin<\rho_l<1$ (wide line) c) Maximal current (MC) phase with bulk density $\pmc=\bar{\rho}$ obtained for $0<\rho_r<\bar{\rho}$ and $\bar{\rho}<\rho_l<1$. Parameter values are $\Delta\mu=20T$, $f=0.5\, f_s$ and the rest as in fig. \ref{fig:pbc}.}
   \label{fig:mch1}
 \end{figure}
We now return to the original problem with open boundary conditions (OBC). The MCP prediction, eq. \eqref{eq:mch}, can be thought of as a variational statement for the bulk density that together with the expression for $\jmpbc$, eq. \eqref{eq:jmpbc}, leads to three qualitatively different phases depending on the relative values of the reservoir densities $\rho_l$ and $\rho_r$, as depicted in fig. \ref{fig:mch1}. Figure  \ref{fig:mch1}a illustrates the so-called low-density (LD) phase that occurs when $\rho_l$ is smaller than the density $\bar{\rho}=\text{argmax}_{\rho}\jmpbc(\rho)$ that maximizes $\jmpbc$, and when $\rho_r$ is smaller than $\prmax$, where $\prmax$ fulfills $\jmpbc(\rho_l)=\jmpbc(\prmax)$. The bulk density attains the value $\pld=\rho_l$ in this case, since the current takes the value $\jm=\jmpbc(\rho_l)$, as predicted by eq. \eqref{eq:mch}. On the other hand, when $\rho_r$ is larger than $\bar{\rho}$, and $\rho_l$ is larger than the critical value $\plmin$ determined by $\jmpbc(\rho_r)=\jmpbc(\plmin)$, the system is in the high-density (HD) phase, and the bulk density is $\phd=\rho_r$, since the current takes the value $\jm=\jmpbc(\rho_r)$ according to eq. \eqref{eq:mch}, see fig. \ref{fig:mch1}b. Finally, for relatively low right reservoir densities, $\rho_r<\bar{\rho}$, and relatively high left reservoir densities, $\rho_l>\bar{\rho}$, the system bulk dynamics becomes independent of the boundary conditions, and the bulk density assumes the maximal current value, $\pmc=\bar{\rho}$.

The reservoir densities can now be calculated in terms of the transition rates by employing current conservation. Equating the bulk current in the LD phase with the current at the left boundary yields
\begin{equation}
  \label{eq:pl}
 \jmpbc(\rho_l)=\alpha(1-\rho_l), 
\end{equation}
which can be solved analytically for $\rho_l$ in terms of the attachment rate $\alpha$ and the single-motor jumping rates $\omega_{ij}$. Similarly, current conservation in the HD phase determines the right reservoir density $\rho_r$ as a function of $\omega_{ij}$ and the detachment rate $\beta$ through
\begin{equation}
  \label{eq:pr}
 \jmpbc(\rho_r)=\beta \rho_r, 
\end{equation}
where we assume that track detachment in model IIa can proceed from any mechanochemical state.

\begin{figure} 
 \psfrag{beta}[ct][ct][1.]{$\beta$ $(\si{\second^{-1}})$}
 \psfrag{alpha}[ct][ct][1.]{$\alpha$ $(\si{\second^{-1}})$}
 \psfrag{LD}[ct][ct][1.]{\textcolor{red}{LD}}
 \psfrag{HD}[ct][ct][1.]{\textcolor{red}{HD}}
 \psfrag{MC}[ct][ct][1.]{\textcolor{red}{MC}}
 \psfrag{legendlegend1}[lt][lt][1.]{$f=0$}
 \psfrag{legendlegend2}[lt][lt][1.]{$f=0.5 f_s$}
 \psfrag{legendlegend3}[lt][lt][1.]{$f=0.9 f_s$}
\centering
\includegraphics[width=.9\columnwidth]{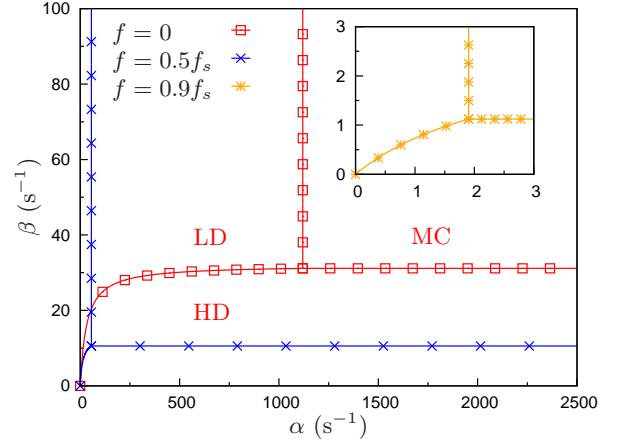}
\caption{Model IIa (OBC): Mean-field phase diagram in the attachment and detachment rates $\alpha$ and $\beta$, respectively, predicted by the MCP for $\Delta\mu=20T$ and $f=0,0.5\,f_s$ (main figure) and $f=0.9\,f_s$ (inset). Labelling of the phases by LD, HD and MC applies to the $f=0$ phase diagram (squares). Legends and parameters are as in fig. \ref{fig:pbc}.}
   \label{fig:mch2}
 \end{figure}
The procedure sketched in figure \ref{fig:mch1} together with eqs.~\eqref{eq:pl}--\eqref{eq:pr} allow us to compute analytically the complete mean-field phase diagram for any choice of the parameter set. The phase diagram projected onto the attachment rate $\alpha$ and the detachment rate $\beta$ is shown in fig. \ref{fig:mch2} for $\Delta\mu=20T$ and for three different values of the load force.
The transition between the LD and the MC phase takes place at $\alpha=\alpha_c$, where $\alpha_c$ fulfills $\rho_l(\alpha_c)=\bar{\rho}$, cf. fig.~\ref{fig:mch1}a and \ref{fig:mch1}c. The phase boundary between the HD and the MC phases occurs at $\beta=\beta_c$, where $\beta_c$ is given by $\rho_r(\beta_c)=\bar{\rho}$, see fig. \ref{fig:mch1}b-c.
Finally, the LD-HD boundary is characterized by $\jmpbc(\rho_l(\alpha))=\jmpbc(\rho_r(\beta))$ as indicated in fig. \ref{fig:mch1}a-b. We note that the MC phase grows with increased load in agreement with the phase diagram obtained for model I, see fig.~\ref{fig:phasediagI}. 
Furthermore, the phase diagram is generally asymmetric in $\alpha$ and $\beta$ due to the lack of particle-hole symmetry, and the LD-HD boundary is highly non-linear for small to intermediate load forces. The non-linearity arises from the interplay of the ASEP exclusion rule, imposed on the mechanical transitions, with the internal chemical transitions that are not affected by steric interactions. The velocity and the hydrolysis rate are simply determined from the phase diagram as $\vobc=\vpbc(\rho_i)$ and $\robc=\rpbc(\rho_i)$, respectively, where $i$ is either LD, HD or MC.

Finally, it is interesting to note that by using the MCP together with the same arguments discussed in this section, one finds that for any exclusion process with internal dynamics whose current $\jmpbc$ in PBC has a single maximum as the one depicted in fig. \ref{fig:mch1}, the phase diagram is qualitatively identical to the one presented in fig. \ref{fig:mch2} with three different  phases.

\subsubsection{Strong exclusion}
In the following we consider the consequences of strong exclusion for many-motor kinetics. In this scenario the ASEP exclusion rule affects \emph{all} the transitions between internal states, and not only the stepping transitions where spatial displacement occurs, as illustrated in fig. \ref{fig:modelIIa}c. The mean-field steady-state master equations for PBC thus become
\begin{align}
\begin{split}
  0=& - \omega_{k\, k-1}\rho^k(1-\rho)+\omega_{k-1 \, k}\rho^{k-1}(1-\rho) \\
                & +\omega_{k \, k+1}\rho^k(1-\rho)-\omega_{k+1 \, k}\rho^{k+1}(1-\rho),  
\end{split}
\label{eq:MFIIastrong}
\end{align}
with $k=1,\dots,6$, and can easily be solved to obtain 
\begin{equation}
  \label{eq:jmpbcIIastrong}
  \jmpbc=v_0 \rho (1-\rho)/a
\end{equation}
for the mechanical probability current. Here, $v_0$ is the single motor velocity of non-interacting motors as obtained from the six-state model \cite{Liepelt2007}. Similarly, the chemical current can be written as
\begin{equation}
  \label{eq:jcpbcIIastrong}
  \jcpbc=r_0 \rho (1-\rho),
\end{equation}
where $r_0$ is the ATP hydrolysis rate of non-interacting motors. By comparing eq. \eqref{eq:jmpbcIIastrong} with the corresponding eq. \eqref{eq:jpbcI} for model I it is tempting to believe that strong exclusion gives rise to qualitatively similar thermodynamics of interacting motors as the tightly coupled model I. However, there is an important difference between the two cases. For model I, the coupling ratio $v/r a$ is by construction equal to the corresponding ratio $v_0/r_0 a=1$ for non-interacting motors. While eqs. \eqref{eq:jmpbcIIastrong}--\eqref{eq:jcpbcIIastrong} imply $v/a r=v_0/a r_0$ in the presence of strong exclusion, a single motor, as described by model IIa, is in general not tightly coupled, i.e. $v_0/r_0 a <1$. Hence, when chemical transitions are directly affected by steric exclusion, only futile hydrolysis events due to traffic jams are prevented, and the coupling ratio becomes independent of motor density.
It is also worth noting that strong exclusion restores the particle-hole symmetry in model IIa. As a consequence, the density in the MC phase is $\bar{\rho}=1/2$, and the phase diagram becomes symmetric, with $\alpha_c=\beta_c$, as already discussed for model I.  

\subsection{Model IIb (PBC-LK)}
\label{sec:modelIIb}
\begin{figure}
   \centering
   \includegraphics[width=\columnwidth]{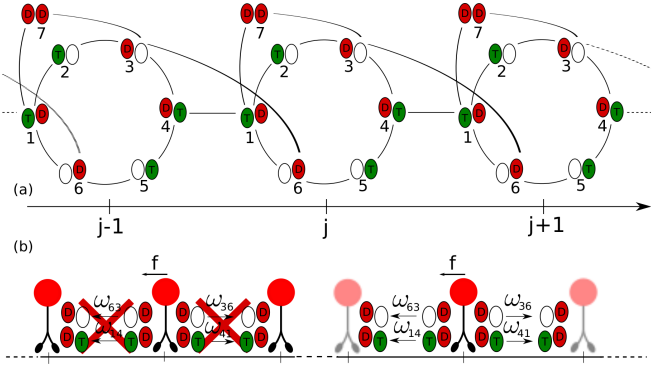}
   \caption{Model IIb: a)  The extended seven-state network model for single kinesin with two mechanical stepping transitions that is presented in \cite{Liepelt2010} and used to model the experimental data of Carter and Cross \cite{Carter2005}. b) The standard ASEP exclusion rule prevents the motor from performing the two stepping transitions between states 1 and 4 and 3 and 6, respectively, if the neighbouring site is occupied.} 
   \label{fig:modelIIb}
 \end{figure}
We now turn to the seven-state model for single kinesin molecules operating under external load that is introduced by Liepelt and Lipowsky in \cite{Liepelt2010}. This model extends the previously considered six-state model \cite{Liepelt2007} by introducing an additional motor state, state 7, and an additional stepping transition, see fig. \ref{fig:modelIIb}a. Since both motor heads are loosely bound to the filament in the newly introduced state, the motor is most likely to detach from the track from state 7. Hence, the seven-state model admits a precise description of the detachment pathway, which is necessary for considering Langmuir-type kinetics. The master equations  for model IIb with Langmuir kinetics are discussed in App.~\ref{sec:app3}. In order to solve these equations under steady state conditions, one has to resort to numerical techniques. 

Furthermore, we consider the effect of strong exclusion (SE) on the dynamics within this model. In analogy to strong exclusion dynamics in model IIa, cf. eqs. \eqref{eq:jmpbcIIastrong}--\eqref{eq:jcpbcIIastrong}, the mechanical and chemical probability currents are given by  
\begin{align}
\begin{split}
\jmlk&=v_0 \plk (1-\plk)/a, \\
\jclk&=r_0 \plk (1-\plk),
\end{split}
\label{eq:jmjcIIb}
\end{align}
where the Langmuir density $\plk$ is obtained as
\begin{equation}
  \label{eq:plkIIb}
  \plk=\frac{\alpha}{\alpha+\beta \rho^7_0}.
\end{equation}
Here, $\alpha$ is the rate with which free motors bind to an empty lattice site, while $\beta$ is the bulk detachment rate characterizing the unbinding of motors from an occupied lattice site. The quantity $\rho^7_0$ is the probability to find a non-interacting motor in the detachment state 7, while $v_0$ and $r_0$ denote, respectively, the motor velocity and hydrolysis rate of non-interacting motors calculated from the seven-state model. The motor velocity and hydrolysis rate of single motors in the presence of interactions are thus linear functions of the motor density.
\begin{align}
\begin{split}
 \vlk&=v_0(1-\plk), \\
 \rlk&=r_0(1-\plk).
\end{split}
\label{eq:vrLKIIb}
\end{align}

\begin{figure} 
 \psfrag{legendlegend1}[lt][lt][1.]{$f=0$, ME}
 \psfrag{legendlegend2}[lt][lt][1.]{$f\neq 0$, ME}
 \psfrag{legendlegend3}[lt][lt][1.]{$f=0$, SE}
 \psfrag{legendlegend4}[lt][lt][1.]{$f\neq 0$, SE}
 \psfrag{ylabelvar}[ct][ct][1.]{$\vlk/a$, $\, \rlk$ (1/s)}
 \psfrag{abyb}[ct][ct][1.]{$\alpha/\beta_0$}
 \psfrag{jlk}[ct][ct][1.]{$\jmlk$, $\jclk$ (1/s)}
 \psfrag{rho}[ct][ct][1.]{$\plk$}
 \psfrag{a}[cB][cB][1.]{{\scriptsize $\textbf{(a)}$}}
 \psfrag{b}[cB][cB][1.]{{\scriptsize $\textbf{(b)}$}}
 \psfrag{c}[cB][cB][1.]{{\scriptsize $\textbf{(c)}$}}
\centering
\includegraphics[width=\columnwidth]{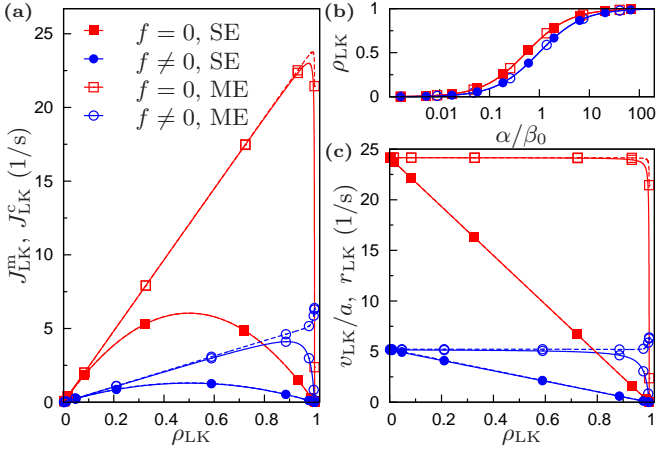}
\caption{Model IIb (PBC-LK): a) A parametric plot of the mechanical current $\jmlk$ (solid lines) and the chemical current $\jclk$ (dashed lines) as a function of the motor density $\plk$ for $\Delta\mu=20T$ and for two different values of the load force, $f=0$ (squares, red) and $f=\SI{-3}{\pico\newton}$ (circles, blue). The empty symbols correspond to the mechanical exclusion (ME) rule inspired by the standard ASEP, while the filled symbols represent the strong exclusion (SE) for all transitions. b) The Langmuir density $\plk$ as a function of the ratio $\alpha/\beta_0$. c) The velocity $\vlk$ and the hydrolysis rate $\rlk$ as a function of $\plk$ obtained from a) and b). Parameter values are those obtained in \cite{Liepelt2010} for the Carter and Cross experiment \cite{Carter2005}.}
   \label{fig:IIbLK}
 \end{figure}
In fig. \ref{fig:IIbLK}a we report the probability currents $\jmlk$ and $\jclk$ as a function of the Langmuir density $\plk$ obtained by varying the ratio $\alpha/\beta_0$ of the binding rate to the (zero-force) unbinding rate. The curves correspond to the free energy $\Delta\mu=20T$ and to two different values of the external load force, $f=0$ and $f=\SI{-3}{\pico\newton}$. Furthermore, we plot the numeric results obtained for the mechanical exclusion (ME) rule as dictated by the standard ASEP as well as the curves for the strong exclusion (SE) rule, eq. \eqref{eq:jmjcIIb}. We note that the motor stall force for ME is a function of the rates $\alpha$ and $\beta$, and as such varies with $\plk$. The density $\plk$, which is obtained numerically for ME and is given by eq. \eqref{eq:plkIIb} for SE, is in general a fast increasing function of $\alpha/\beta_0$ as shown in fig. \ref{fig:IIbLK}b. Interestingly, $\plk$ is essentially the same for standard and strong exclusion, even though the probability currents are largely affected when adding exclusion on chemical transitions. Finally, fig. \ref{fig:IIbLK}c shows the resulting motor velocity $\vlk=\jmlk a/\plk$ and hydrolysis rate $\rlk=\jclk/\plk$. For ME (curves with empty symbols in the figure), the velocity is essentially constant and equal to the velocity of non-interacting motors, i.e. $\vlk \simeq v_0$, for a wide range of densities, $0<\plk \lesssim 0.9$, which is in agreement with the experimental observations of Seitz and Surrey on the kinetics of conventional kinesin (kinesin-1) motor traffic at zero force \cite{Seitz2006}. 
It is also worth noting that our result for the velocity in the presence of Langmuir kinetics is at variance with that of Klumpp, Chai and Lipowsky \cite{Klumpp2008a}: by using a simplified two state model, they conclude that the internal cycle has a small effect on the linear decrease of velocity with increasing density, such a decrease being expected for systems with no internal dynamics, cf. eq.~\eqref{eq:vlkI}. 
However, it is interesting to note that the linearly decreasing density-velocity relationship, eq. \eqref{eq:vrLKIIb}, is recovered for SE, i.e. when the exclusion rule affects both chemical and mechanical transitions. As opposed to Seitz and Surrey \cite{Seitz2006}, in a recent experiment Leduc et al. \cite{Leduc2012} observe a linear relationship between motor density and velocity for kinesin-8 motors moving on crowded filaments. Hence, our modelling suggests that the contradictory experimental results for $v(\rho)$ could be attributed to different extents of strong exclusion that different motor types might exhibit, in the different experiments.

\section{Model II: Maximum power regime}
\label{sec:maxpowerII}
In the following we investigate the operation of loosely coupled molecular machines in the maximum power regime. Model IIa with six internal states is used to investigate systems where the particle exchange predominantly occurs at the lattice boundaries, while the seven-state model IIb allows us to consider motors interacting with a bulk reservoir.

\subsection{Efficiency at maximum power}
\label{sec:EMPII}
\begin{figure}
 \psfrag{etas}[ct][ct][1.]{$\etaobc^*$}
 \psfrag{Dmu}[ct][ct][1.]{$\Delta\mu/T$}
 \psfrag{Ia}[lB][lB][1.]{{\scriptsize  $\textbf{(Ia)}$ }}
 \psfrag{Ib}[lB][lB][1.]{{\scriptsize  $\textbf{(Ib)}$ }}
 \psfrag{IIa}[lB][lB][1.]{{\scriptsize $\textbf{(IIa)}$}}
 \psfrag{IIb}[lB][lB][1.]{{\scriptsize $\textbf{(IIb)}$}}
 \psfrag{phil1}[lB][lB][1.]{{\scriptsize  $\phi=0$}}
 \psfrag{phil2}[lB][lB][1.]{{\scriptsize  $\phi=0.1$}}
 \psfrag{chi1l1}[lB][lB][1.]{{\scriptsize  $\chi_1=0.15$}}
 \psfrag{chi1l2}[lB][lB][1.]{{\scriptsize  $\chi_1=0.3$}}
 \psfrag{chi2l1}[lB][lB][1.]{{\scriptsize  $\chi_2=0.25$}}
 \psfrag{chi2l2}[lB][lB][1.]{{\scriptsize  $\chi_2=0.4$}}
 \psfrag{legendlegend1}[rc][rc][1.]{{\scriptsize non-int}}
 \psfrag{legendlegend2}[rc][rc][1.]{{\scriptsize $\alpha=5$, ME}}
 \psfrag{legendlegend3}[rc][rc][1.]{{\scriptsize $\alpha=5$, SE}}
 \psfrag{legendlegend4}[rc][rc][1.]{{\scriptsize $\alpha=50$, ME}}
 \psfrag{legendlegend5}[rc][rc][1.]{{\scriptsize $\alpha=50$, SE}}
\centering
\includegraphics[width=\columnwidth]{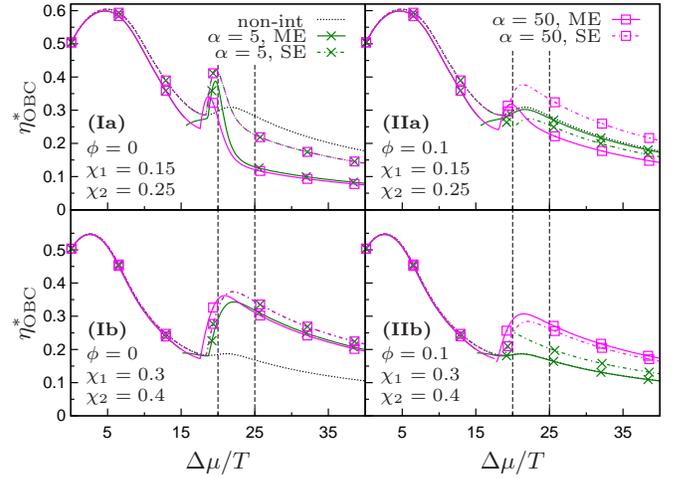}
\caption{Model IIa (OBC):  EMP as a function of the free energy $\Delta\mu/T$ for mechanical exclusion (solid lines) and for strong mechanochemical exclusion (dot-dashed lines) for two different attachment rates, $\alpha=\SI{5}{\second^{-1}}$ (green crosses) and $\alpha=\SI{50}{\second^{-1}}$ (magenta squares). First row (a): Parameter values as obtained in \cite{Liepelt2007} for the Carter and Cross experiment \cite{Carter2005}, second row (b): same parameters as in (I) but with the chemical load factors $\chi_1=0.3$ and $\chi_2=0.4$ taken from the seven-state model \cite{Liepelt2010} (see text). The detachment rate force sensitivity is $\phi=0$ (I) and $\phi=0.1$ (II), respectively. The value of the zero-force unbinding rate is $\beta_0=\SI{3}{\second^{-1}}$. The biologically relevant regime $\Delta\mu/T \sim 20-25$ \cite{Howard2001} is marked by vertical dashed lines.}
   \label{fig:modelIIaemp}
 \end{figure}
The optimal force $f^*$ maximizing the output power for model II is calculated as described in sec. \ref{sec:maxpowerI} for model I. Since the mechanical and chemical currents are not tightly coupled, the EMP for a fixed value of $\Delta\mu$ is then obtained as
\begin{equation}
  \eta^*=-\frac{f^* v^*}{\Delta\mu r^*},
\end{equation}
where $r^*=r(f^*)$ is the hydrolysis rate at maximum power. We find that the EMP is only susceptible to variations of a few parameters, namely the binding rate $\alpha$, the unbinding force dependence parameter $\phi$, and the chemical load factors $\chi_1$ and $\chi_2$ that describe the load dependence of the transition rates associated to chemical transitions \footnote{See \cite{Golubeva2012a,Liepelt2007} for a definition and discussion of $\chi_1$ and $\chi_2$}. There is in general no consensus in literature on the biologically applicable values for these parameters since they are highly model dependent and/or experiment dependent as will be discussed below. In fig. \ref{fig:modelIIaemp} we plot the EMP for a few combinations of parameter values for $\alpha$, $\phi$ and $\chi_i$. The values are chosen such as to show the typical qualitative behaviour for the EMP when the parameters are varied within a range compatible with experimental results reported in literature. We therefore believe that the results for the EMP shown in fig. \ref{fig:modelIIaemp} are representative for the rather broad range of biologically relevant parameter values.

\begin{figure}
 \psfrag{MC}[ct][ct][1.]{{\scriptsize MC}}
 \psfrag{HD}[ct][ct][1.]{{\scriptsize HD}}
 \psfrag{LD}[ct][ct][1.]{{\scriptsize LD}}
 \psfrag{Dmu1}[lt][lt][1.]{{\scriptsize $\Delta\mu=15.3T$}}
 \psfrag{Dmu2}[lt][lt][1.]{{\scriptsize $\Delta\mu=17T$}}
 \psfrag{Dmu3}[ct][ct][1.]{{\scriptsize $\Delta\mu=19T$}}
 \psfrag{Dmu4}[ct][ct][1.]{{\scriptsize $\Delta\mu=22T$}}
 \psfrag{pout}[ct][ct][1.]{$\Pout$ (pN nm/s)}
 \psfrag{f}[ct][ct][1.]{$f$ (pN)}
 \psfrag{a}[cB][cB][1.]{{\scriptsize $\textbf{(a)}$}}
 \psfrag{b}[cB][cB][1.]{{\scriptsize $\textbf{(b)}$}}
 \psfrag{c}[cB][cB][1.]{{\scriptsize $\textbf{(c)}$}}
 \psfrag{d}[cB][cB][1.]{{\scriptsize $\textbf{(d)}$}}
\centering
\includegraphics[width=\columnwidth]{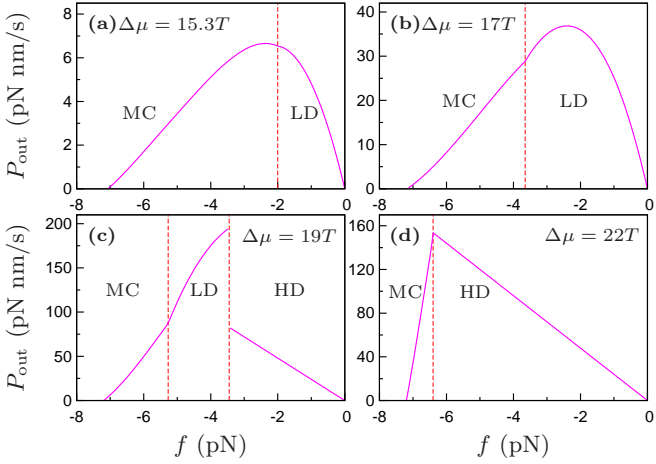}
\caption{Model IIa (OBC): The output power $\Pout$ as a function of the external load force $f$ for different values of the chemical free energy, $\Delta\mu=15.3T$ (a),$\Delta\mu=17T$ (b), $\Delta\mu=19T$ (c) and $\Delta\mu=22T$ (d). Parameters are as in (Ia) in fig. \ref{fig:modelIIaemp} with $\alpha=\SI{5}{\second^{-1}}$. We consider the ME rule here. The vertical dashed (red) lines indicate phase transitions.}
   \label{fig:modelIIapout}
 \end{figure}

\begin{figure}
 \psfrag{MC}[ct][ct][1.]{{\scriptsize MC}}
 \psfrag{HD}[ct][ct][1.]{{\scriptsize HD}}
 \psfrag{LD}[ct][ct][1.]{{\scriptsize LD}}
 \psfrag{label1}[ct][ct][1.]{{\scriptsize $\alpha=5,\phi=0$}}
 \psfrag{label2}[ct][ct][1.]{{\scriptsize $\alpha=50,\phi=0$}}
 \psfrag{label3}[ct][ct][1.]{{\scriptsize $\alpha=5,\phi=0.1$}}
 \psfrag{eta}[ct][ct][1.]{$\eta$}
 \psfrag{pout}[ct][ct][1.]{$\Pout$ (pN nm/s)}
 \psfrag{f}[ct][ct][1.]{$f$ (pN)}
 \psfrag{la}[ct][ct][1.]{{\footnotesize \textcolor{green}{$\eta_0^*$}}}
 \psfrag{lb}[cc][cc][1.]{{\footnotesize \textcolor{green}{$\eta^*$}}}
 \psfrag{Ia}[lB][lB][1.]{{\scriptsize $\textbf{(Ia)}$}}
 \psfrag{Ib}[lB][lB][1.]{{\scriptsize $\textbf{(Ib)}$}}
 \psfrag{Ic}[lB][lB][1.]{{\scriptsize $\textbf{(Ic)}$}}
 \psfrag{IIa}[rB][rB][1.]{{\scriptsize $\textbf{(IIa)}$}}
 \psfrag{IIb}[rB][rB][1.]{{\scriptsize $\textbf{(IIb)}$}}
 \psfrag{IIc}[rB][rB][1.]{{\scriptsize $\textbf{(IIc)}$}}
\centering
\includegraphics[width=\columnwidth]{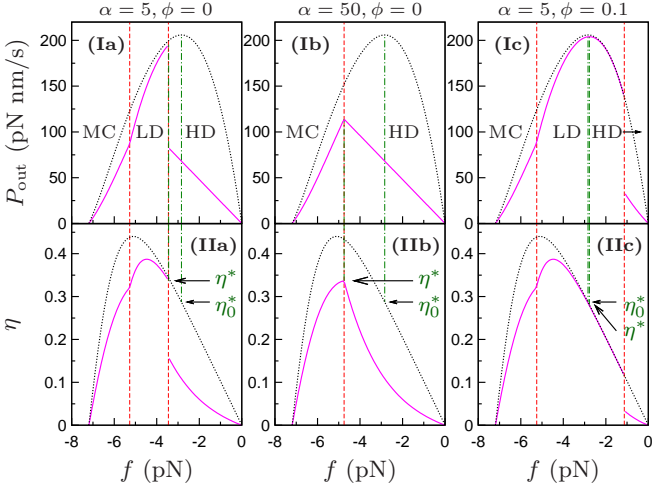}
\caption{Model IIa (OBC): The output power $\Pout$ (row I) and the efficiency $\eta$ (row II) as a function of the external load force $f$ for $\Delta\mu=19T$ and for different parameter values: $\alpha=\SI{5}{\second^{-1}}$ and $\phi=0$ (a), $\alpha=\SI{50}{\second^{-1}}$ and $\phi=0$ (b), and $\alpha=\SI{5}{\second^{-1}}$ and $\phi=0.1$ (c). The rest of the parameters are as in fig. \ref{fig:modelIIaemp} (row I). We consider the ME rule here.}
   \label{fig:modelIIapout2}
 \end{figure}

The solid lines in fig. \ref{fig:modelIIaemp}(Ia) show the EMP predicted for model IIa with mechanical exclusion calculated using the parameter values obtained in \cite{Liepelt2007} by fitting the Carter and Cross' experimental data \cite{Carter2005}  and for force independent detachment dynamics, i.e.  $\phi=0$. 
For both $\alpha=\SI{5}{\second^{-1}}$ (green crosses) and $\alpha=\SI{50}{\second^{-1}}$ (magenta squares) the single-motor EMP exhibits an enhancement as compared to noninteracting motor systems under the same conditions when the (dimensionless) free energy $\Delta\mu/T$ lies in the range $18-22$. 
Thus, the region of EMP enhancement partly overlaps with the biologically compatible range $\Delta\mu/T \sim 20-25$ \cite{Howard2001}. As for model I, the EMP increase is induced by the presence of phase transitions in the collective dynamics of interacting motors as depicted in fig. \ref{fig:modelIIapout} for a specific set of parameters. Here, we plot the output power $\Pout$ as a function of the load force $f$ for various values of $\Delta\mu$. For small values of $\Delta\mu$ the maximum power output is reached in the MC phase, and the EMP $\eta^*$ is essentially equal to the EMP $\eta^*_0$ of noninteracting motors, see fig. \ref{fig:modelIIapout}a. For $15.7 \lesssim \Delta\mu/T \lesssim 18.6$ the motors operate at maximum power in the LD phase, see fig. \ref{fig:modelIIapout}b, and $\eta^*<\eta^*_0$ since the velocity $v$ in the LD phase decreases more rapidly with $f$ than the velocity $v_0$ of non-interacting motors. When $\Delta \mu/T$ increases beyond the value $18.6$, the maximum of $\Pout$ lies at the boundary between the HD and LD phase as illustrated in fig. \ref{fig:modelIIapout}c. The optimal force $f^*$ is thus greater than $f^*_0$, and the EMP $\eta^*$ exceeds the non-interacting value $\eta^*_0$ as shown in fig. \ref{fig:modelIIapout2}a. At $\Delta\mu/T \sim 22$ the force $f^*$ becomes so large that $\eta^*=\eta^*_0$. For even larger values of $\Delta\mu$, the maximizing force $f^*$ approaches the stall force, and $\eta^*\to 0$ due to dissipation. Furthermore, for $\Delta\mu/T \gtrsim 22$ the maximum power occurs at the MC-HD boundary, see fig. \ref{fig:modelIIapout}d.

For a larger value of the binding rate, $\alpha=\SI{50}{\second^{-1}}$, the EMP also exhibits an enhancement, which is, however, smaller than for $\alpha=\SI{5}{\second^{-1}}$. We also note that the maximum of the output power lies at the MC-HD boundary in the region of enhancement as shown in  \ref{fig:modelIIapout2}b. We stress, nevertheless, that the physical mechanism leading to the EMP increase is identical for the two situations, namely that motor-motor interactions affect the characteristic response of the velocity to external mechanical driving. For  $\alpha=\SI{5}{\second^{-1}}$ the EMP boost disappears rapidly with increasing values of $\phi$, see fig. \ref{fig:modelIIaemp}(IIa), because the maximum power is obtained in the LD regime as pictured in fig. \ref{fig:modelIIapout2}c. The EMP curves for $\alpha=\SI{50}{\second^{-1}}$ are more robust to changes in $\phi$, since the motors operate deeper in the HD phase for $f=0$, see fig. \ref{fig:mch2}. This effect is enhanced by the fact that the HD-LD boundary is highly non-linear as discussed in sec. \ref{sec:MCP}. 

Together with the EMP results obtained for the ME rule and discussed above, we plot in fig. \ref{fig:modelIIaemp} the EMP predicted for the SE case (dot-dashed lines). We observe that the EMP increase is present and pronounced for both values of $\alpha$ when $\phi=0$, fig. \ref{fig:modelIIaemp}(Ia). However, even for SE, the effect disappears for $\alpha=\SI{5}{\second^{-1}}$ when increasing $\phi$ for the same reasons as for ME. On the other hand, for $\alpha=\SI{50}{\second^{-1}}$ the EMP increase is robust towards variations in $\phi$, because the maximum of the output power lies at the MC-HD transition as for ME. Hence, interestingly, for small $\alpha$ the  dependence of the EMP on $\phi$ is determined by the intrinsic dissipation present in the single-motor model rather than by traffic jam related futile hydrolysis.

In figs. \ref{fig:modelIIaemp}(Ib)-(IIb) we report the EMP where the chemical parameters take on the values obtained by fitting the seven-state model \cite{Liepelt2010} to the experimental data of Carter and Cross, while all the other model parameters are kept constant. For this parameter set we observe an approximately twofold increase in the EMP as compared to the non-interacting case within a wide range of free energies, $\Delta\mu/T\sim 18-40$, and for all combinations of parameter values and exclusion types except one. However, the absence of EMP enhancement for the case  $\alpha=\SI{5}{\second^{-1}}$, $\phi=0.1$, with mechanical exclusion seems to be due to the fact that we have considered the simplified six-state model: indeed, by considering the same parameter values for the seven-state model, we find an increase in the EMP, see App. \ref{sec:app4}. It is worth to note that, for negligible values of $\phi$, the EMP for interacting molecular machines is essentially independent of the binding rate and type of exclusion considered. 

A discussion of the parameter values used in the present study is now in order. 

By comparing the values for the chemical load parameters for the six-state model, $\chi_1=0.15$ and $\chi_2=0.25$, with the corresponding values for the seven-state model, $\chi_1=0.3$ and $\chi_2=0.4$, we conclude that they depend strongly on the underlying model used to fit the experimental data. Furthermore, the parameters $\chi_i$ assume different values when the six-state model is used to fit two different sets of experimental data \cite{Liepelt2007}. Thus, the values for $\chi_i$ are also sensitive to experimental conditions.
 Lastly, it is worth noting that the parameters for model II have been obtained in refs. \cite{Liepelt2007,Liepelt2010} by fitting experimental data for single motor movement, and thus it is also possible that the parameters $\chi_i$ assume different values in the presence of multiple motors on the track. This conjecture is based on the fact that the description of the dynamics of a single motor as a stochastic hopping process between discrete mechanochemical states is merely an approximation to the microscopic picture where the motor movement is represented as Brownian diffusion on a continuous free energy landscape \cite{Keller2000,Magnasco1994,Golubeva2012}. From this point of view, the states of the discrete model correspond to the potential minima of the energy landscape, and the transition rates of the discrete model are obtained from the minimal energy paths connecting the potential minima. Since motor-motor interactions alter the energy landscape, it is thus likely that the parameters representing the coupling of the external force to the mechanochemistry will change as a consequence. In ref.~\cite{Golubeva2012}, for example, we address the question of how the load factor $\theta$ introduced in model~I changes with the external force, in a system with continuous phase space.

To our best knowledge, there are very few precise measurements of $\phi$ available in literature. The value $\phi=0.1$ for the force dependence of the unbinding rate is obtained in ref.~\cite{Liepelt2007} by fitting a simplified seven-state model with only one mechanical transition to the Visscher et al. experiment \cite{Visscher1999}. Since the parameters $\chi_i$ describing the effect of the external load force on chemical transitions are both model dependent and experiment dependent, it is hence likely that $\phi$ would be so as well and thus attain a value different from $0.1$ when the full seven-state model is fitted to the experimental data of Carter and Cross. 

The value of the binding rate $\alpha$ is widely assumed to depend linearly on the (local) motor concentration \cite{Klumpp2008a,Nishinari2005}. Since intracellular kinesin concentrations vary over several orders of magnitude \cite{Howard2001}, there is therefore reason to believe that $\alpha$ attains a wide range of values in biological systems. For example, the values used for $\alpha$ in the present paper lie within the biological range $0.8-80 \si{\second^{-1}}$ obtained from the binding constant $\SI{8.27E7}{\Molar^{-1}\second^{-1}}$ \cite{Grant2011} and intracellular kinesin concentrations $(10-1000) \, \cdot 10^{-9} \si{\Molar}$ \cite{Nishinari2005}. 

In summary, considering the results for the EMP presented in this section and the above discussion on parameter uncertainty, we believe that the EMP enhancement observed takes place for biological systems under a wide range of biological conditions. 

\begin{figure}
 \psfrag{etas}[ct][ct][1.]{$\etalk^*$}
 \psfrag{Dmu}[ct][ct][1.]{$\Delta\mu/T$}
 \psfrag{legendlegendlegend1}[lc][lc][1.]{{\scriptsize non-int}}
 \psfrag{legendlegendlegend2}[lc][lc][1.]{{\scriptsize $\alpha=5$, ME}}
 \psfrag{legendlegendlegend3}[lc][lc][1.]{{\scriptsize $\alpha=5$, SE}}
 \psfrag{legendlegendlegend4}[lc][lc][1.]{{\scriptsize $\alpha=50$, ME}}
 \psfrag{legendlegendlegend5}[lc][lc][1.]{{\scriptsize $\alpha=50$, SE}} 
 \psfrag{a}[lB][lB][1.]{{\scriptsize $\textbf{(a)}$ $\phi=0$}}
 \psfrag{b}[lB][lB][1.]{{\scriptsize $\textbf{(b)}$ $\phi=0.1$}}
\centering
\includegraphics[width=\columnwidth]{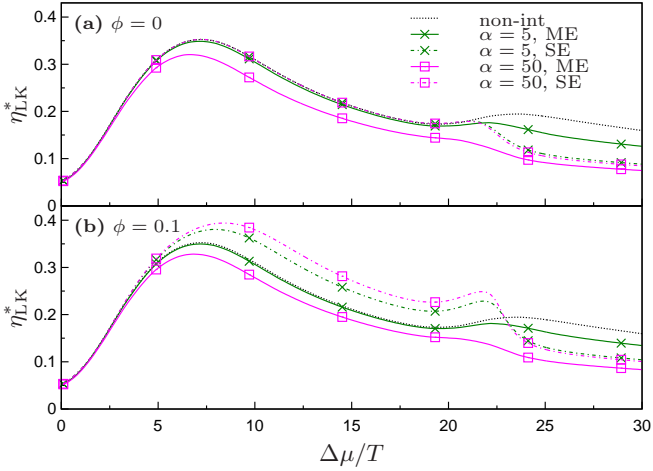}
\caption{Model IIb (PBC-LK):  EMP as a function of $\Delta\mu/T$ for mechanical ASEP exclusion (solid lines) and for strong mechanochemical exclusion (dot-dashed lines) for two different attachment rates, $\alpha=\SI{5}{\second^{-1}}$ (green crosses) and $\alpha=\SI{50}{\second^{-1}}$ (magenta squares), and for two different values of $\phi$: $\phi=0$ (a) and $\phi=0.1$ (b). Parameter values are as obtained in \cite{Liepelt2010}, and $\beta_0=\SI{3}{\second^{-1}}$.}
   \label{fig:modelIIbemp}
 \end{figure}

Finally, we consider the EMP for model IIb which accounts for the behaviour of the efficiency when particle exchange predominantly occurs in the bulk. As for model IIa, we investigate the dependence of the EMP on $\alpha$, $\phi$ and exclusion type and plot typical curves in fig. \ref{fig:modelIIbemp}. We note, however, that, as opposed to the system with OBC, the EMP calculated within model IIb changes in a continuous manner when the parameters are varied since no dynamical phase transitions are present in the case of Langmuir kinetics. For $\phi=0$, see fig. \ref{fig:modelIIbemp}a, the EMP in the presence of motor-motor interactions is smaller or approximately equal to the EMP of non-interacting motors for a wide range of $\alpha$-values and for both mechanical and strong exclusion. The EMP is generally decreasing with increasing $\alpha$. The EMP for $\phi=0.1$ is shown in fig. \ref{fig:modelIIbemp}b. In the presence of exclusion on both mechanical and chemical transitions (SE), the EMP exhibits a considerable increase as compared to $\eta^*_0$ in the region $\Delta\mu/T\sim 5-22$. Furthermore, the EMP increases with $\alpha$ for these values of the free energy as opposed to the observed general tendency. Interestingly, as opposed to model IIa, the motor density variations induced by imposing steric exclusion on mechanical stepping only are not sufficient to observe EMP enhancement in the case of LK.

\subsection{Velocity and density at maximum power}
\label{sec:velocity_densityII}
\begin{figure}
 \psfrag{vs}[ct][ct][1.]{$\vobc^*$ (nm/s)}
 \psfrag{rhos}[ct][ct][1.]{$\pobc^*$}
 \psfrag{Dmu}[ct][ct][1.]{$\Delta\mu/T$}
 \psfrag{Ia}[lB][lB][1.]{{\scriptsize  $\textbf{(Ia)}$ }}
 \psfrag{Ib}[lB][lB][1.]{{\scriptsize  $\textbf{(Ib)}$ }}
 \psfrag{IIa}[lB][lB][1.]{{\scriptsize $\textbf{(IIa)}$}}
 \psfrag{IIb}[lB][lB][1.]{{\scriptsize $\textbf{(IIb)}$}}
 \psfrag{phil1}[lB][lB][1.]{{\scriptsize  $\phi=0$}}
 \psfrag{phil2}[lB][lB][1.]{{\scriptsize  $\phi=0.1$}}
 \psfrag{chi1l2}[lB][lB][1.]{{\scriptsize  $\chi_1=0.3$}}
 \psfrag{chi2l2}[lB][lB][1.]{{\scriptsize  $\chi_2=0.4$}}
 \psfrag{legendlegend1}[lc][lc][1.]{{\scriptsize $\alpha=5$, ME}}
 \psfrag{legendlegend2}[lc][lc][1.]{{\scriptsize $\alpha=5$, SE}}
 \psfrag{legendlegend3}[lc][lc][1.]{{\scriptsize $\alpha=50$, ME}}
 \psfrag{legendlegend4}[lc][lc][1.]{{\scriptsize $\alpha=50$, SE}}
\centering
\includegraphics[width=\columnwidth]{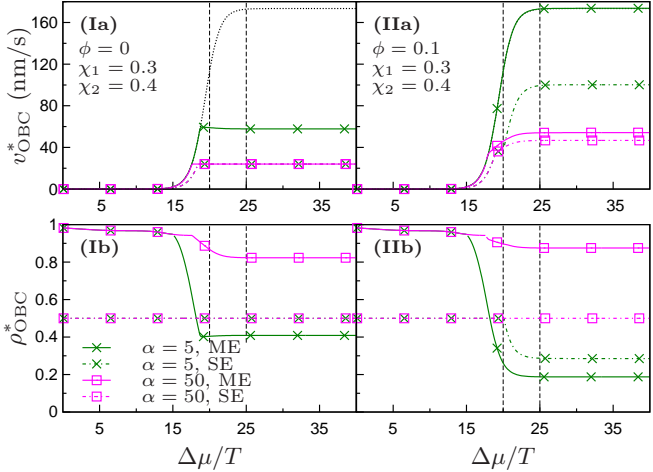}
\caption{Model IIa (OBC):  Velocity at maximum power $\vobc^*$ (a) and density at maximum power $\pobc^*$ (b) as a function of the free energy $\Delta\mu/T$ for two different values of $\phi$: $\phi=0$ (I) and $\phi=0.1$ (II). Legends and parameters are as in fig. \ref{fig:modelIIaemp}b.}
   \label{fig:modelIIavrho}
 \end{figure}
We proceed by discussing the VMP and DMP of interacting kinesin motors as obtained within model II. Fig. \ref{fig:modelIIavrho} shows the VMP $v^*$ (a) and DMP $\rho^*$ (b) for the parameters used to calculate the EMP in fig. \ref{fig:modelIIaemp}b. The corresponding results for the second set of parameters used in fig. \ref{fig:modelIIaemp}a are similar and therefore not presented here. First of all, we note that, contrary to the behaviour predicted by the tightly-coupled model I, see fig. \ref{fig:modelIvstar}b, $v^*$ at most differs from the VMP of non-interacting motors $v^*_0$ by one order of magnitude. This effect is due to the saturation of the velocity at biological values of the chemical input as discussed in App.~\ref{sec:app1}. Secondly, the VMP is generally lowered when the exclusion rule affects both the mechanical and chemical transitions. Furthermore, in analogy to model I, the VMP and EMP exhibit a trade-off behaviour; the EMP decreases for increasing values of $\phi$, while the VMP increases. As for model I, the DMP never exceeds $0.5$ for SE, since maximum power is never achieved in the high-density regime. For the ME rule, however, the DMP reaches $\rho^*\sim 0.85 $ for $\alpha=\SI{50}{\second^{-1}}$ since the density in the MC phase is shifted to higher values, see fig. \ref{fig:mch1}. 
Moreover, the DMP is sensitive to the value of $\alpha$.

\begin{figure}
 \psfrag{vs}[ct][ct][1.]{$\vlk^*$ (nm/s)}
 \psfrag{rhos}[ct][ct][1.]{$\plk^*$}
 \psfrag{Dmu}[ct][ct][1.]{$\Delta\mu/T$}
 \psfrag{Ia}[lB][lB][1.]{{\scriptsize  $\textbf{(a)}$ }}
 \psfrag{Ib}[lB][lB][1.]{{\scriptsize  $\textbf{(b)}$ }}
 \psfrag{phil1}[lB][lB][1.]{{\scriptsize  $\phi=0.1$}}
 \psfrag{legendlegend1}[lc][lc][1.]{{\scriptsize non-int}}
 \psfrag{legendlegend2}[lc][lc][1.]{{\scriptsize $\alpha=5$, ME}}
 \psfrag{legendlegend3}[lc][lc][1.]{{\scriptsize $\alpha=5$, SE}}
 \psfrag{legendlegend4}[lc][lc][1.]{{\scriptsize $\alpha=50$, ME}}
 \psfrag{legendlegend5}[lc][lc][1.]{{\scriptsize $\alpha=50$, SE}}
\centering
\includegraphics[width=\columnwidth]{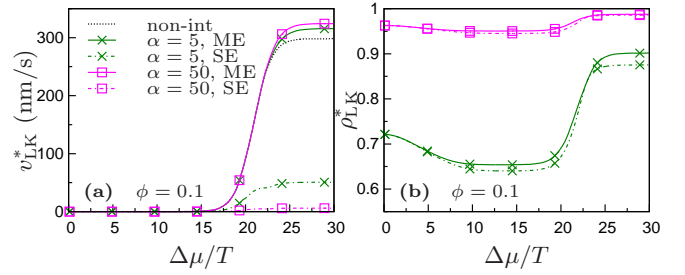}
\caption{Model IIb (PBC-LK):  Velocity at maximum power $\vlk^*$ (a) and density at maximum power $\plk^*$ (b) as a function of the chemical input $\Delta\mu/T$ for $\phi=0.1$. Legends and parameters are as in fig. \ref{fig:modelIIbemp}b.}
   \label{fig:modelIIbvrho}
 \end{figure}

Finally, in fig. \ref{fig:modelIIbvrho} we report the VMP (a) and the DMP (b) for model IIb. Interestingly, for ME, the VMP can exceed the corresponding VMP $v^*_0$ in non-interacting systems for high values of $\Delta\mu$. However, the high VMP comes at the expense of the EMP $\eta^*$ as can be seen in fig. \ref{fig:modelIIbemp}b. For strong mechanical and chemical exclusion, where an EMP enhancement is observed for a wide range of $\Delta\mu$-values, the velocity is reduced in analogy to model IIa. In analogy to model I, the DMP does not vary significantly with $\Delta\mu$. Contrary to model IIa, the DMP is essentially unaffected by chemical exclusion and is above $0.5$ for all values of $\Delta\mu$.

\section{Discussion and conclusions}
\label{sec:conclusion}
In the present paper we have considered two different models for kinesin motors interacting through steric exclusion on a filamentous track. Furthermore, we have investigated the effect of different binding and unbinding kinetics on the overall dynamics of the interacting system as well as on the operation in the maximum power regime. In the following we discuss the relevance of our findings to intracellular transport and in the context of state-of-the-art multiple-motor experiments. 

In the present work we have for simplicity been concerned with single-motor-single-cargo systems. There is indeed experimental evidence that justifies this assumption. It has been found in several \emph{in vivo} studies on different motor types (see \cite{Ross2008} and references therein) as well as in a recent \emph{in vitro} experiment on kinesin-1 \cite{Jamison2010}, that only a single motor is actively engaged at any one time in carrying its cargo. Furthermore, another experimental study \cite{Driver2011} performed with multiple kinesin-1 motors concludes that cooperative effects for this motor are small, albeit they increase with efficiency. Nonetheless, it is widely believed, primarily based on results from \emph{in vitro} experiments, that multiple motors are involved in cargo transport in the cell giving rise to non-steric coupling between the motors. A recent state-of-the-art \emph{in vivo} study of intracellular transport of lipid droplets in Drosophila embryos \cite{Leidel2012} corroborates this conclusion. The typical run length of a single kinesin motor is known to be around $\SI{1}{\micro\meter}$ as reported in, e.g., \cite{Liepelt2007,Klumpp2008a,Seitz2006}. Since microtubule lengths are at least one order of magnitude larger than the single-motor run length \cite{Howard2001}, one would therefore expect the Langmuir dynamics to be non-negligible when the cargo is transported by a single motor. However, several experimental as well as theoretical works report that the cargo run length increases by up to several orders of magnitude when multiple motors carry a single cargo \cite{Furuta2013,Beeg2008,Derr2012,Klumpp2005}. Even under the assumption that only one motor is active at any time, the presence of other motors on the cargo would likely lead to an enhancement in the run length. Hence, under physiological conditions one would expect the cargo run lengths to be comparable to the microtubule lengths. Furthermore, the density of the cargo is not homogeneous across the cell, so one would  rather expect that the cargo concentration gradient led to the cargo binding primarily occurring at the filament end closer to the production site. Based on the above arguments, we thus believe that intracellular traffic is best described by a model with open boundary conditions. 

In that context we find that the EMP of kinesin motors is increased, as compared to the non-interacting case, due to a change in the characteristic force-velocity relation, $v(f)$, as discussed in sec. \ref{sec:EMPII}. Remarkably, for the two kinesin models studied, it occurs in the biologically relevant portion of the parameter space and for different types of exclusion rules. Still, the results have been obtained for the single-motor-single-cargo dynamics, and future developments of the present research would be concerned with intracellular transport involving several motors on a cargo. However, we  believe that our conclusions can be carried over to the many-motor-single-cargo system, as discussed below. When a cargo is transported by multiple motors, a question arises regarding the sharing of the load force amongst the motors, as also discussed in \cite{Guerin2010}. Leidel et al. \cite{Leidel2012} arrive at the conclusion that the load force is shared equally by the motors, and that only one type of motors is involved in pulling the droplets at any one time, i.e. there is no tug-of-war mechanism as has been suggested by several authors, see, e.g., ref.~\cite{Mueller2008} and references therein. It is certainly true that the presence of multiple motors on a cargo would modify $v(f)$, as also discussed in \cite{Guerin2010,Klumpp2005,Campas2006,Campas2008,Brugues2009}. 
However, based on the assumption of homogeneous load-sharing and identical motors discussed above, we believe that $v(f)$ would only change quantitatively, and not qualitatively, see, e.g., \cite{Klumpp2005}, showing thus a monotonic behaviour as for the case we consider here. Hence, our single-motor-single-cargo model system can be thought of as a renormalized version of the multiple-motor-single-cargo system. One may therefore expect the multiple-motor-single-cargo systems to undergo a dynamic phase transition from a high-density phase to a phase with a lower density, resulting in an enhancement in the EMP, similar to the one we find for our models.

Despite the recent tremendous progress achieved within experimental tracking and manipulating techniques, it is as yet not possible to study under controlled conditions the dynamics of multiple motors operating under externally applied load forces. One class of state-of-the-art motor traffic experiments deals with traffic of single motors under zero-load conditions \cite{Seitz2006,Leduc2012}. In this case, the density of the free unbound motors surrounding the filament is expected to be homogeneous across the system, and the run length is short compared to the filament length, as discussed above. Under such conditions we therefore expect the Langmuir dynamics to dominate. Furthermore, it has been proposed that the use of particular tracking techniques such as quantum-dot labelling can influence motor-motor interactions \cite{Telley2009}. Hence, the extent of exclusion can well depend on the experimental setup, which the different density-velocity relations, $v(\rho)$, could be a sign of, see sec. \ref{sec:modelIIb}. We expect that future motor traffic experiments would provide a better estimation for the model parameters which have only been extracted from experimental data on single motors so far.

Finally, we note that the enhancement in EMP is present for each variant of model kinesin we considered, for a wide choice of parameters: this shows that this effect does not originate from the details of the microscopic dynamics, but is rather a result of the cooperative nature of the system. We therefore believe that the interaction-caused enhancement of the efficiency due to a change in the force-velocity response is a generic feature for nanomachines optimized for transport in the maximum power regime. 

\acknowledgements
We gratefully acknowledge financial support from Lundbeck Fonden.

\appendix 

\section{Model I. Parameter estimation}
\label{sec:app1}
\begin{figure}
 \psfrag{Dmu}[ct][ct][1.]{$\Delta\mu/T$}
 \psfrag{v}[ct][ct][1.]{$\vsI(f=0),\vsII(f=0)$ (nm/s) }
 \psfrag{legendlegend1}[lt][lt][1.]{Carter \& Cross \cite{Carter2005}}
 \psfrag{legendlegend2}[lt][lt][1.]{Visscher et al. \cite{Visscher1999}}
\centering
\includegraphics[width=\columnwidth]{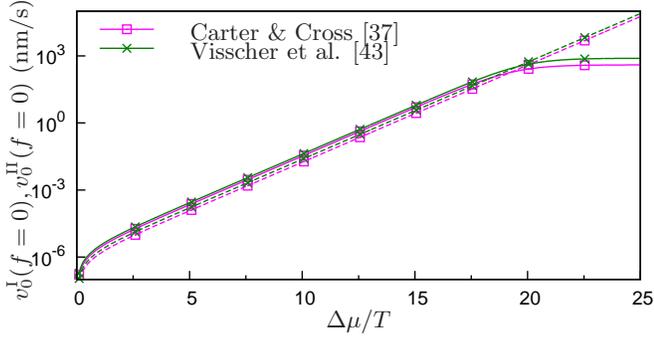}
\caption{The single motor velocity at zero load force, $\vsII(f=0)$, obtained in the absence of interactions for model II (solid lines) together with the best fit, $\vsI(f=0)$, obtained within model I (dashed lines). The two different parameter sets correspond to the values obtained by Liepelt and Lipowsky in \cite{Liepelt2007} for the Carter and Cross experiment \cite{Carter2005} (magenta, squares) and for the Visscher et al. experiment \cite{Visscher1999} (green, crosses), respectively. The corresponding values of the microscopic rate $\omega_0$ that yield the best fit of the velocity are $\omega_0=\SI{1.00E-7}{\second^{-1}}$ and $\omega_0=\SI{1.33E-7}{\second^{-1}}$, respectively, for the two experiments.}
   \label{fig:modelIfit}
 \end{figure}

The 6-state model discussed in \cite{Liepelt2007} and which model IIa is based on provides a good description of the experimental data obtained in several single-molecule experiments on kinesin \cite{Liepelt2007}. We therefore derive the parameter values for model I from the motor kinetic properties as obtained in model IIa in the following way. The microscopic transition rate $\omega_0$ is fixed by comparing the single particle velocity in model I with the corresponding velocity for model IIa. To be more precise, we plot for two different parameter sets for model IIa the single motor velocity in the absence of interactions and at zero load force, $\vsII(f=0)$, as a function of the chemical input $\Delta\mu$. Fitting the resulting curves by the expression 
\begin{equation}
\vsI(f=0)=a \omega_0 (e^{\Delta\mu/T}-1)  
\end{equation}
for model I then yields the estimate of $\omega_0$ for model I, see fig. \ref{fig:modelIfit}. For parameters obtained by fitting the experimental data of Carter and Cross \cite{Carter2005}, we obtain $\omega_0=\SI{1.00E-7}{\second^{-1}}$, while using the parameters obtained by Visscher et al.'s data \cite{Visscher1999} leads to $\omega_0=\SI{1.33E-7}{\second^{-1}}$.  It is worth noting, however, that energy dissipation due to loose coupling results in saturation of the velocity at $\Delta\mu \gtrsim 20 k_BT$, see fig. \ref{fig:modelIfit}, which is not present in the tightly coupled model I. 

For the attachment rate $\alpha$ and the bare detachment rate $\beta_0$ we take $\SI{5}{\second^{-1}}$ \cite{Klumpp2005} and $\SI{3}{\second^{-1}}$ \cite{Liepelt2007}, respectively. For the force dependence of the detachment rate Liepelt and Lipowsky find $\phi=0.1$ \cite{Liepelt2007}. However, since the EMP strongly depends on $\phi$, cf. fig. \ref{fig:modelIemp}, we consider different values of $\phi$ in sec. \ref{sec:maxpowerI}. The mechanical load sharing parameter $\theta$ is sensitive to the choice of model and to experimental conditions \cite{Liepelt2007,Liepelt2010}. In model I we use the values $\theta=0.65$ and $\theta=0.3$ as obtained in \cite{Liepelt2007} for the Carter and Cross and for the Visscher et al. experiment, respectively.

\section{Model I. Critical values $\Delta\mu_{c,1}$ and $\Delta\mu_{c,2}$}
\label{sec:app2}
The value of $\Delta\mu_{c,1}$ that separates the regimes of maximum power in the MC phase and at the HD-MC phase transition, respectively, is determined in the following way. The location of the maximum in the MC phase, $\fobc^*(\Delta\mu)$, is obtained by solving (cf. eq. \eqref{eq:vOBCI})
\begin{equation}
  \der{\Pout}{f}=\der{}{f} \big( - \frac{a}{2} f (p(\Delta\mu,f)-q(f)) \big) =0.
\end{equation}
Equating the velocity at maximum power in the MC phase with the velocity in the HD phase then yields a transcendental relation for $\Delta\mu_{c,1}$,
\begin{equation}
  v_{\text{MC}}(f^*(\Delta\mu_{c,1}),\Delta\mu_{c,1})=v_{\text{HD}}(f^*(\Delta\mu_{c,1})).
\end{equation}
The relationship between the thermodynamic and mechanical quantities as quantified in the parametrization of the jumping rates $p$ and $q$, eq. \eqref{eq:rates}, thus leads to a weak, logarithmic dependence of $\Delta\mu_{c,1}$ on the model parameters.

For the quantity $\Delta\mu_{c,2}$, that is meaningful whenever $\phi \neq 0$, a closed form expression can be obtained. The condition $\alpha=\beta$ for the LD-HD boundary yields the transition force
\begin{equation}
  \fhdld=-\log(\alpha\beta_0)T/\phi a,
  \label{eq:fhdld}
\end{equation}
while the critical value of the input free energy occurs when the point $(\alpha,\beta)$ coincides with the triple point $(\alpha_c,\beta_c)$ for $f=\fhdld$. The equation for $\Delta\mu_{c,2}$ thus reads
\begin{equation*}
  \alpha=\beta=\beta_c=\left( p(\Delta\mu_{c,2},\fhdld)-q(\Delta\mu_{c,2},\fhdld) \right)/2,
\end{equation*}
and can be solved to obtain
\begin{equation*}
  \Delta\mu_{c,2}(\theta,\phi)=T\log(2\alpha/\omega_0+e^{-\fhdld a(1-\theta)/T})-\fhdld a\theta.
\end{equation*}
Note that $\Delta\mu_{c,2}$ is a function of $\phi$ due to the dependence of $\fhdld$ on $\phi$, eq. \eqref{eq:fhdld}.

\section{Model IIb. Master equations}
\label{sec:app3}
The mean-field master equations describing the steady-state dynamics on a periodic lattice in the presence of Langmuir kinetics are
\begin{align}
\begin{split}
  0=&(\omega_{41}\rho^4-\omega_{14}\rho^1) (1-\rho)+\omega_{71}\rho^7-\omega_{17}\rho_1  \\
                &+ \omega_{21}\rho^2-\omega_{12}\rho^1+\omega_{61}\rho^6-\omega_{16}\rho^1   \\ 
  0=&-\omega_{21}\rho^2+\omega_{12}\rho^1 + \omega_{32}\rho^3-\omega_{23}\rho^2    \\
  0=&(\omega_{63}\rho^6-\omega_{36}\rho^3)(1-\rho)+\omega_{73}\rho^7-\omega_{37}\rho^3 \\
               &- \omega_{32}\rho^3+\omega_{23}\rho^2+ \omega_{43}\rho^4-\omega_{34}\rho^3 \\
  0=& -(\omega_{41}\rho^4 - \omega_{14}\rho^1)(1-\rho)  \\
                  & -\omega_{43}\rho^4+\omega_{34}\rho^3+\omega_{54}\rho^5-\omega_{45}\rho^4     \\
  0=& -(\omega_{63}\rho^6-\omega_{36}\rho^3)(1-\rho) \\
                  & -\omega_{54}\rho^5+\omega_{45}\rho^4+ \omega_{65}\rho^6-\omega_{56}\rho^5  \\
  0=& - \omega_{65}\rho^6+\omega_{56}\rho^5 +\omega_{61}\rho^6-\omega_{16}\rho^1  \\
  0=&  -\omega_{73}\rho^7+\omega_{37}\rho^3 -\omega_{71}\rho^7+\omega_{17}\rho_1 \\
                 &+\alpha(1-\rho)-\beta \rho^7, 
\end{split}
\label{eq:MFmodelIIb}
\end{align}
since motors detach from the filament from state 7. Furthermore, we have assumed that attachment likewise proceeds through state 7. If binding and unbinding take place from different internal states, the chemical potential differences involved in the two processes are different. Hence, thermodynamic consistency requires that an additional term due to particle exchange with the reservoir is introduced in the input power. Since experimental values for these chemical potential differences are not available at present, we have chosen to minimize the number of unknown parameters by taking state 7 as the binding state.  

\section{Model IIb. OBC}
\label{sec:app4}
\begin{figure}[h]
 \psfrag{etas}[ct][ct][1.]{$\etaobc^*$}
 \psfrag{Dmu}[ct][ct][1.]{$\Delta\mu/T$}
 \psfrag{legendlegend1}[lc][lc][1.]{{\scriptsize non-int}}
 \psfrag{legendlegend2}[lc][lc][1.]{{\scriptsize $\phi=0.1$, $\alpha=5$}}
\centering
\includegraphics[width=\columnwidth]{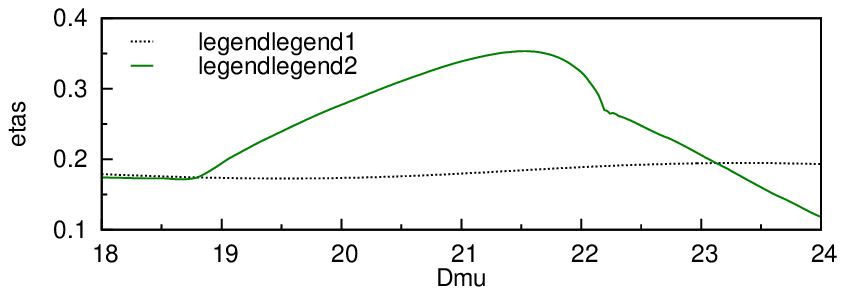}
\caption{Model IIb (OBC):  Parameter values: $\alpha=\SI{5}{\second^{-1}}$, $\beta_0=\SI{3}{\second^{-1}}$, $\phi=0.1$. }
   \label{fig:modelIIbOBCemp}
 \end{figure}

The mean-field equations for the seven-state model IIb with OBC can be solved numerically with the help of the MCP. The solution is, however, difficult to obtain due to numerical errors arising as a consequence of the exponential dependencies of the transition rates on the free energy $\Delta\mu$ and load force $f$. However, we have calculated the EMP with mechanical exclusion for $\alpha=\SI{5}{\second^{-1}}$, $\phi=0.1$ and for a limited range of $\Delta\mu$ values, see fig. \ref{fig:modelIIbOBCemp}. As opposed to what we find for the six-state model, fig. \ref{fig:modelIIaemp}(IIb), the resulting EMP exhibits a significant increase, as compared to the non-interacting system, for $\Delta\mu/T\sim 19-23$.

%

\end{document}